\documentclass[twocolumn,prb,showpacs,amsmath,amssymb]{revtex4}
\pdfoutput=1

\usepackage{graphicx}
\usepackage{dcolumn}
\usepackage{bm}
\usepackage[pdftex]{color}

\begin{document}

\title{Quantum fluctuations in the effective pseudospin-$1/2$ model \\ for magnetic pyrochlore oxides}

\author{Shigeki Onoda}
\author{Yoichi Tanaka}
\affiliation{%
Condensed Matter Theory Laboratory, RIKEN, 2-1, Hirosawa, Wako 351-0198, Saitama, JAPAN
}%

\date{\today}

\begin{abstract}
The effective quantum pseudospin-$1/2$ model for interacting rare-earth magnetic moments, which are locally described with atomic doublets, is studied theoretically for magnetic pyrochlore oxides. It is derived microscopically for localized Pr$^{3+}$ $4f$ moments in Pr$_2TM_2$O$_7$ ($TM=$Zr, Sn, Hf, and Ir) by starting from the atomic non-Kramers magnetic doublets and performing the strong-coupling perturbation expansion of the virtual electron transfer between the Pr $4f$ and O $2p$ electrons. The most generic form of the nearest-neighbor anisotropic superexchange pseudospin-$1/2$ Hamiltonian is also constructed from the symmetry properties, which is applicable to Kramers ions Nd$^{3+}$, Sm$^{3+}$, and Yb$^{3+}$ potentially showing large quantum effects. The effective model is then studied by means of a classical mean-field theory and the exact diagonalization on a single tetrahedron and on a 16-site cluster. These calculations reveal appreciable quantum fluctuations leading to quantum phase transitions to a quadrupolar state as a melting of spin ice for the Pr$^{3+}$ case. The model also shows a formation of cooperative quadrupole moment and pseudospin chirality on tetrahedrons. A sign of a singlet quantum spin ice is also found in a finite region in the space of coupling constants. The relevance to the experiments is discussed.
\end{abstract}

\maketitle

\section{Introduction}
\label{sec:intro}

Quantum fluctuations and geometrical frustration are a couple of key ingredients in realizing nontrivial spin-disordered ground states without a magnetic dipole long-range order (LRO) in three spatial dimensions~\cite{anderson:56,anderson,wen:89,palee:09}. The pyrochlore lattice structure is a typical example where the geometrical frustration plays a crucial role in preventing the LRO~\cite{anderson:56,reimers:91,moessner:98}. Of our particular interest in this paper is a so-called dipolar spin ice~\cite{harris:97,ramirez:99,bramwell:01,gardner:10}, such as Dy$_2$Ti$_2$O$_7$ and Ho$_2$Ti$_2$O$_7$, and related systems. The dipolar spin ice provides a classical magnetic analogue of a cubic water ice~\cite{pauling} and is characterized by the emergent U(1) gauge field mediating the Coulomb interaction between monopole charges~\cite{hermele:04,castelnovo:08} as well as the dipolar spin correlation showing a pinch-point singularity~\cite{isakov:04,henley:05}. Introducing quantum effects to the classical spin ice may produce further nontrivial states of matter. Evidence of quantum effects has recently been observed with inelastic neutron-scattering experiments on the spin-ice related compounds, Tb$_2$Ti$_2$O$_7$~\cite{gardner:99,gardner:01,mirebeau:07}, Tb$_2$Sn$_2$O$_7$~\cite{mirebeau:07}, and Pr$_2$Sn$_2$O$_7$~\cite{zhou:08}. Exploiting a weak coupling of the rare-earth magnetic moments to conduction electrons, a chiral spin state~\cite{wen:89} has been detected through the anomalous Hall effect~\cite{ahe} at zero magnetic field without magnetic dipole LRO in another related compound Pr$_2$Ir$_2$O$_7$~\cite{machida:09}. Vital roles of the planar components have also been experimentally observed in Yb$_2$Ti$_2$O$_7$ and Er$_2$Ti$_2$O$_7$~\cite{cao:09}. Obviously, quantum fluctuations enrich the otherwise classical properties of the spin ice. They may drive it to other states of matter, including quadrupolar states and chiral spin states~\cite{onoda:09}. The aims of this paper are to provide a comprehensive derivation of a realistic effective quantum model for these spin-ice related materials and to understand its basic properties including nontrivial quantum effects.

\subsection{Classical dipolar spin ice}

Let us briefly review the classical (spin) ice. The low-energy properties of water and spin ices are described by Ising degrees of freedom that represent whether proton displacements (electric dipoles) and magnetic dipoles, respectively, point inwards (``in'') to or outwards (``out'') from the center of the tetrahedron. The interaction among the Ising variables favors nearest-neighbor pairs of ``in'' and ``out'' and thus suffers from geometrical frustration. This produces a so-called ice rule~\cite{bernal:33,pauling} stabilizing ``2-in, 2-out'' configurations on each tetrahedron. Macroscopic degeneracy of this ice-rule manifold produces Pauling's residual entropy $\frac{1}{2}R\ln\frac{3}{2}$~\cite{pauling}. 

In the dipolar spin ice~\cite{harris:97,ramirez:99,bramwell:01}, a rare-earth magnetic moment $\hat{\bm{m}}_{\bm{r}}=g_J \mu_B \hat{\bm{J}}_{\bm{r}}$ located at a vertex $\bm{r}$ of tetrahedrons plays the role of the Ising variable because of the large crystalline electric field (CEF), which is often {\it approximately} modeled by
\begin{equation}
  \hat{H}_{\mathrm{Ising}}=-D_{\mathrm{Ising}}\sum_{\bm{r}}(\bm{n}_{\bm{r}}\cdot\hat{\bm{J}}_{\bm{r}}/J)^2,
  \label{eq:H_Delta}
\end{equation}
with the Land\'{e} factor $g_J$ and $D_{\mathrm{Ising}}>0$. Here, $J$ is the quantum number for the total angular momentum $\hat{\bm{J}}_{\bm{r}}$, and $\bm{n}_{\bm{r}}$ defines a unit vector at a pyrochlore-lattice site $\bm{r}$ that points outwards from the center of the tetrahedron belonging to one fcc sublattice of the diamond lattice and inwards to that belonging to the other sublattice.
The amplitude of the rare-earth magnetic moment is so large that the interaction between the magnetic moments is dominated by the magnetic dipolar interaction~\cite{rossat-mignod:83,hertog:00}, 
\begin{equation}
  \hat{H}_{\mathrm{D}}=\frac{\mu_0}{4\pi}\sum_{\langle\bm{r},\bm{r}'\rangle}\left[\frac{\hat{\bm{m}}_{\bm{r}}\cdot\hat{\bm{m}}_{\bm{r}'}}{({\mit\Delta}r)^3}-3\frac{(\hat{\bm{m}}_{\bm{r}}\cdot{\mit\Delta}\bm{r})({\mit\Delta}\bm{r}\cdot\hat{\bm{m}}_{\bm{r}'})}{({\mit\Delta}r)^5}\right],
  \label{eq:H_D}
\end{equation}
with ${\mit\Delta}\bm{r}=\bm{r}-\bm{r}'$ and the summation $\sum_{\langle\bm{r},\bm{r}'\rangle}$ over all the pairs of atomic sites. 
This yields a ferromagnetic coupling $D_{\mathrm{n.n.}}=\frac{5}{3}\frac{\mu_0}{4\pi}\frac{m^2}{(a/2\sqrt{2})^3}\sim2.4$~K between the nearest-neighbor magnetic moments for Ho$_2$Ti$_2$O$_7$ and Dy$_2$Ti$_2$O$_7$ with the moment amplitude $m\sim10\mu_B$ and the lattice constant $a\sim10.1$~\AA~\cite{bramwell:01}, providing a main driving force of the ice rule.
It prevails over the nearest-neighbor superexchange interaction which is usually {\it assumed} to take the isotropic Heisenberg form
\begin{equation}
  \hat{H}_{\mathrm{H}}=-3J_{\mathrm{n.n.}}\sum_{\langle\bm{r},\bm{r}'\rangle}^{\mathrm{n.n.}}\hat{\bm{J}}_{\bm{r}}\cdot\hat{\bm{J}}_{\bm{r}'}/J^2.
  \label{eq:H_H}
\end{equation}
In the limit of $D_{\mathrm{Ising}}\to\infty$,  $\hat{H}_{\mathrm{DSI}}=\hat{H}_{\mathrm{Ising}}+\hat{H}_{\mathrm{D}}+\hat{H}_{\mathrm{H}}$ is reduced to an Ising model~\cite{hertog:00}, which can explain many magnetic properties experimentally observed at temperatures well below the crystal-field excitation energy~\cite{bramwell:01,castelnovo:08,jaubert:09,gardner:10}.

Because of the ferromagnetic effective nearest-neighbor coupling $J_{\mathrm{eff}}=D_{\mathrm{n.n.}}+J_{\mathrm{n.n.}}>0$, creating ``3-in, 1-out'' and ``1-in, 3-out'' configurations out of the macroscopically degenerate ``2-in, 2-out'' spin-ice manifold costs an energy, and can be regarded as defects of magnetic monopoles and anti-monopoles with a unit magnetic charge~\cite{moessner:98}. Then, the spin ice is described as the Coulomb phase of magnetic monopoles where the emergent U(1) gauge fields mediate the Coulomb interaction between monopole charges~\cite{hermele:04,castelnovo:08}. The density of magnetic monopoles is significantly suppressed to lower the total free energy at a temperature $T<J_{\mathrm{eff}}\sim$ a few Kelvin. Simultaneously, the reduction of the monopole density suppresses spin-flip processes, for instance, due to a quantum tunneling~\cite{ehlers:03}, that change the configuration of monopoles. Hence, the relaxation time to reach the thermal equilibrium shows a rapid increase. These phenomena associated with a thermal quench of spin ice have been experimentally observed~\cite{snyder:01,snyder:04} and successfully mimicked by classical Monte-Carlo simulations on the Coulomb gas model of magnetic monopoles~\cite{jaubert:09,castelnovo:10}. This indicates that the quantum effects are almost negligible in the dipolar spin ice.
It has been shown that the emergent gapless U(1) gauge excitations together with a power-law decay of spin correlations can survive against a weak antiferroic exchange interaction that exchanges the nearest-neighbor pseudospin-$1/2$ variables (``in'' and ``out'')~\cite{hermele:04}. This $U(1)$ spin liquid~\cite{hermele:04} can be viewed as a quantum version of the spin ice, though the macroscopic degeneracy of the ice-rule manifold should eventually be lifted in the ideal case under the equilibrium. 

\subsection{Quantum effects}

At a first glance, one might suspect that quantum fluctuations should be significantly suppressed by a large total angular momentum $J$ of the localized rare-earth magnetic moment and its strong single-spin Ising anisotropy $D_{\mathrm{Ising}}>0$, since they favor a large amplitude of the quantum number for $\hat{J}_{\bm{r}}^z\equiv\hat{\bm{J}}_{\bm{r}}\cdot\bm{n}_{\bm{r}}$, either $M_J=J$ or $-J$. Namely, in the effective Hamiltonian, $\hat{H}_{\mathrm{DSI}}=\hat{H}_{\mathrm{Ising}}+\hat{H}_{\mathrm{D}}+\hat{H}_{\mathrm{H}}$, a process for successive flips of the total angular momentum from $M_J=J$ to $-J$ at one site and from $-J$ to $J$ at the adjacent site is considerably suppressed at a temperature $T\ll D_{\mathrm{Ising}}$. The coupling constant for this pseudospin-flip interaction is of order of $|J_{\mathrm{n.n.}}|(|J_{\mathrm{n.n.}}|/D_{\mathrm{Ising}})^{2J}$, and becomes negligibly small compared to the Ising coupling $J_{\mathrm{eff}}$. 

\begin{figure}[tbh]
\begin{center}\leavevmode
\includegraphics[width=\columnwidth]{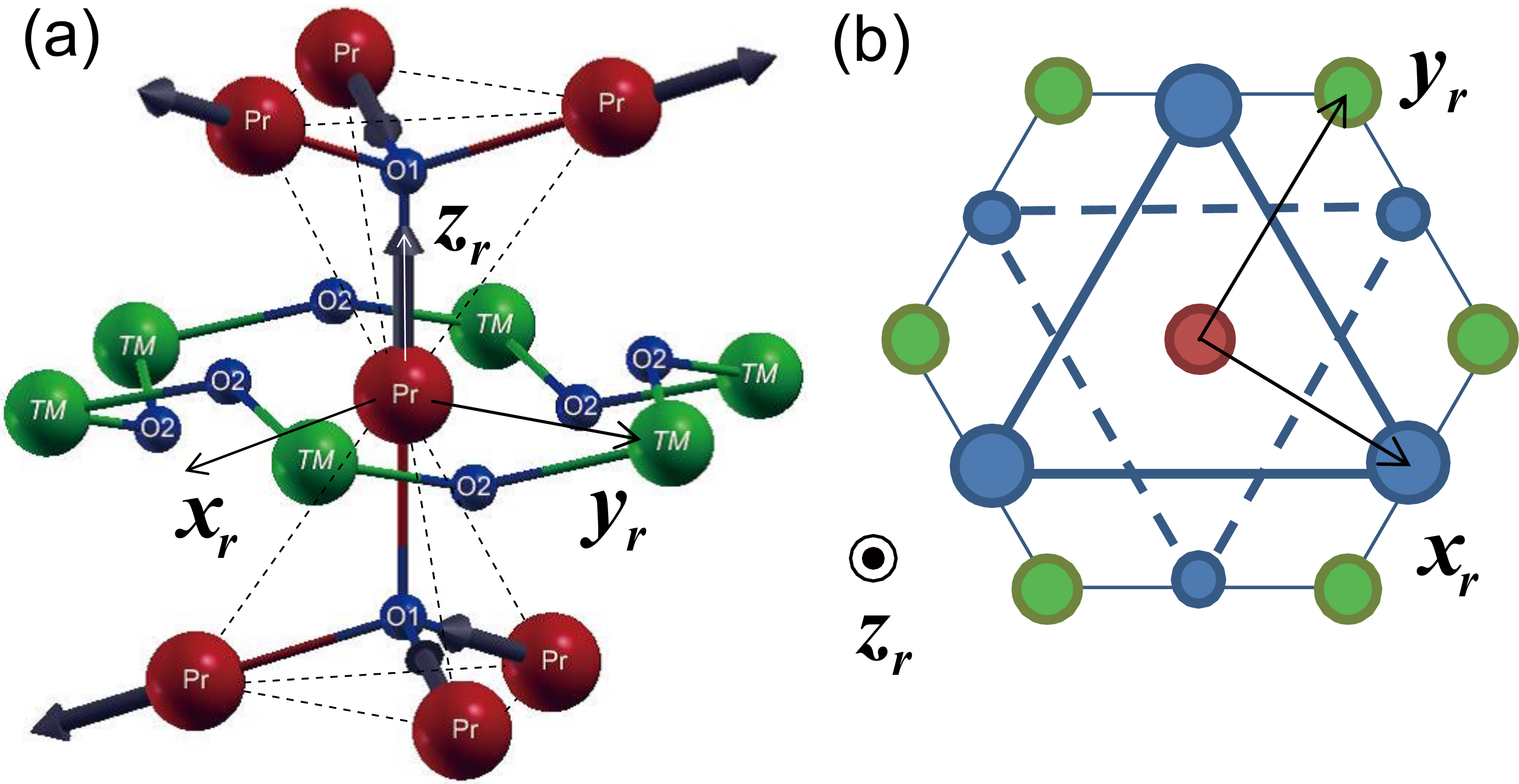}
\end{center}
\caption{(Color online) (a) Pr$^{3+}$ ions form tetrahedrons (dashed lines) centered at O$^{2-}$ ions (O1), and are surrounded by O$^{2-}$ ions (O2) in the $D_{3d}$ symmetry as well as by transition-metal ions ($TM$). Each Pr magnetic moment (bold arrow) points to either of the two neighboring O1 sites. $(\bm{x}_{\bm{r}},\bm{y}_{\bm{r}},\bm{z}_{\bm{r}})$  denotes the local coordinate frame. (b) The local coordinate frame from the top. The upward and the downward triangles of the O$^{2-}$ ions (O2) are located above and below the hexagon of the $TM$ ions. }
\label{fig:crystal}
\end{figure}

In reality, however, because of the $D_{3d}$ crystalline electric field (CEF) acting on rare-earth ions [Fig.~\ref{fig:crystal}], the conservation of $\hat{J}^z$, which is implicitly assumed in the above consideration, no longer holds in the atomic level. Eigenstates of the atomic Hamiltonian including the $LS$ coupling and the CEF take the form of a superposition of eigenstates of $\hat{J}^z$ whose eigenvalues are different by integer multiples of three. Obviously, this is advantageous for the quantum spin exchange to efficiently work. 

Attempts to include quantum effects have recently been in progress.
It has been argued that the presence of a low-energy crystal-field excited doublet above the ground-state doublet in Tb$_2$Ti$_2$O$_7$~\cite{gardner:99,gardner:10} enhances quantum fluctuations and possibly drives the classical spin ice into a quantum spin ice composed of a quantum superposition of ``2-in, 2-out'' configurations~\cite{molavian:07}.

We have proposed theoretically an alternative scenario, namely, a quantum melting of spin ice~\cite{onoda:09}. A quantum entanglement among the degenerate states lifts the macroscopic degeneracy, suppresses the spin-ice freezing, and thus leads to another distinct ground state. Actually, the quantum-mechanical spin-exchange Hamiltonian mixes ``2-in, 2-out'' configurations with ``3-in, 1-out'' and ``1-in, 3-out'', leading to a failure of the strict ice rule and {\it a finite density of monopoles at the quantum-mechanical ground state}. Namely, the quantum-mechanically proliferated monopoles can modify the dipolar spin-ice ground state, while a spatial profile of short-range spin correlations still resembles that of the dipolar spin ice~\cite{onoda:09}. They may appear in bound pairs or in condensates. We have reported that there appears a significantly large anisotropic quantum-mechanical superexchange interaction between Pr magnetic moments in Pr$_2TM_2$O$_7$~\cite{onoda:09} ($TM$ = Zr, Sn, Hf, and Ir)~\cite{subramanian:83}. This anisotropic superexchange interaction drives {\it quantum phase transitions among the spin ice, quadrupolar states having nontrivial chirality correlations, and the quantum spin ice}, as we will see later.

Actually, among the rare-earth ions available for magnetic pyrochlore oxides~\cite{subramanian:83,gardner:10}, the Pr$^{3+}$ ion could optimally exhibit the quantum effects because of the following two facts. 
(i) A relatively small magnitude of the Pr$^{3+}$ localized magnetic moment, whose atomic value is given by $3.2\mu_B$, suppresses the magnetic dipolar interaction, which is proportional to the square of the moment size. Then, for Pr$_2TM_2$O$_7$, one obtains $D_{\mathrm{n.n.}}\sim0.1$~K, which is an order of magnitude smaller than 2.4~K for Ho$_2$Ti$_2$O$_7$ and Dy$_2$Ti$_2$O$_7$. Similarly, quantum effects might appear prominently also for Nd$^{3+}$, Sm$^{3+}$, and Yb$^{3+}$ ions because of their small moment amplitudes, $3.3\mu_B$, $0.7\mu_B$, and $4\mu_B$, respectively, for isolated cases.
(ii) With fewer $4f$ electrons, the $4f$-electron wavefunction becomes less localized at atomic sites. This enhances the overlap with the O $2p$ orbitals at the O1 site [Fig.~\ref{fig:crystal} (a)], and thus the superexchange interaction which is also further increased by a near resonance of Pr $4f$ and O $2p$ levels. Moreover, this superexchange interaction appreciably deviates from the isotropic Heisenberg form because of the highly anisotropic orbital shape of the $f$-electron wavefunction and the strong $LS$ coupling.
Since the direct Coulomb exchange interaction is even negligibly small~\cite{rossat-mignod:83}, this superexchange interaction due to virtual $f$-$p$ electron transfers is expected to be the leading interaction.

Recent experiments on Pr$_2$Sn$_2$O$_7$~\cite{matsuhira:02}, Pr$_2$Zr$_2$O$_7$~\cite{matsuhira:09}, and Pr$_2$Ir$_2$O$_7$~\cite{nakatsuji:06} have shown that the Pr$^{3+}$ ion provides the $\langle111\rangle$ Ising moment described by a non-Kramers magnetic doublet. 
They show similarities to the dipolar spin ice. (i) No magnetic dipole LRO is observed down to a partial spin-freezing temperature $T_f\sim0.1$-$0.3$ K~\cite{matsuhira:02,matsuhira:09,nakatsuji:06,machida:09,zhou:08,maclaughlin:08}. (ii) Pr$_2$Ir$_2$O$_7$ shows a metamagnetic transition at low temperatures only when the magnetic field is applied in the [111] direction~\cite{machida:09}, indicating the ice-rule formation due to the effective ferromagnetic coupling $2J_{\mathrm{eff}}\sim 2J_{\mathrm{n.n.}}\sim1.4$ K~\cite{machida:09}. 
On the other hand, substantially different experimental observations from the dipolar spin ice have also been made. The Curie-Weiss temperature $T_{CW}$ is antiferromagnetic for the zirconate~\cite{matsuhira:09} and iridate~\cite{nakatsuji:06}, unlike the spin ice. The stannate shows a significant level of low-energy short-range spin dynamics in the energy range up to a few Kelvin~\cite{zhou:08}, which is absent in the classical spin ice. Furthermore, the iridate shows the Hall effect at zero magnetic field without magnetic dipole LRO~\cite{machida:09}, suggesting an onset of a chiral spin-liquid phase~\cite{wen:89} at $T_H\sim1.5$~K.

The discovery of this chiral spin state endowed with a broken time-reversal symmetry on a macroscopic scale in Pr$_2$Ir$_2$O$_7$ without apparent magnetic LRO~\cite{machida:09} has increased the variety of spin liquids. One might speculate that this is caused mainly by a Kondo coupling to Ir conduction electrons and thus the RKKY interaction~\cite{rkky}. 
However, the low-temperature thermodynamic properties are common in this series of materials, Pr$_2TM_2$O$_7$, except that a small partial reduction ($\sim10\%$) of Pr magnetic moments probably due to conduction electrons affects the resistivity and the magnetic susceptibility in Pr$_2$Ir$_2$O$_7$~\cite{nakatsuji:06}.
Furthermore, the onset temperature $T_H\sim1.5$~K for the emergent anomalous Hall effect is comparable to the ferromagnetic coupling $2J_{\mathrm{eff}}\sim1.4$~K ~\cite{machida:09}. Therefore, it is natural to expect that a seed of the chiral spin state below $T_H$ exists in the Pr moments interacting through the superexchange interaction and possibly the state is stabilized by the conduction electrons.
Another intriguing observation here is that without appreciable quantum effects, the chiral manifold of classical ice-rule spin configurations~\cite{machida:09} that has been invented to account for the emergent anomalous Hall effect will result from a magnetic dipole LRO or freezing, which is actually absent down to $T_f$. This points to a significant level of nontrivial quantum effects.

In this paper, we develop a realistic effective theory for frustrated magnets Pr$_2TM_2$O$_7$ on the pyrochlore lattice and provide generic implications on quantum effects in spin-ice related materials, giving a comprehensive explanation of our recent Letter~\cite{onoda:09}. In Sec.~\ref{sec:model}, the most generic nearest-neighbor pseudospin-$1/2$ Hamiltonian for interacting magnetic moments on the pyrochlore lattice is derived on a basis of atomic magnetic doublets for both non-Kramers and Kramers ions. In particular, it is microscopically derived from strong-coupling perturbation theory in the Pr$^{3+}$ case. We analyze the model for the non-Kramers case by means of a classical mean-field theory in Sec.~\ref{sec:MFT}, which reveals spin-ice, antiferroquadrupolar, and noncoplanar ferroquadrupolar phases at low temperatures. Then, we perform exact-diagonalization calculations for the quantum pseudospin-$1/2$ case on a single tetrahedron in Sec.~\ref{sec:single} and on the 16-site cube in Sec.~\ref{sec:ED}. We have found within the 16-site cluster calculations a cooperative ferroquadrupolar phase, which is accompanied by crystal symmetry lowering from cubic to tetragonal and can then be categorized into a magnetic analog of a smectic or crystalline phase~\cite{degenne}. This provides a scenario of the quantum melting of spin ice and can explain the experimentally observed magnetic properties, including powder neutron-scattering experiments on Pr$_2$Sn$_2$O$_7$ and the magnetization curve on Pr$_2$Ir$_2$O$_7$. We also reveal a possible source of the time-reversal symmetry breaking observed in Pr$_2$Ir$_2$O$_7$. It takes the form of the solid angle subtended by four pseudospins on a tetrahedron, each of which is composed of the Ising dipole magnetic moment and the planar atomic quadrupole moment, and shows a nontrivial correlation because of a geometrical frustration associated with the fcc sublattice structure.
A possible sign of a singlet quantum spin-ice state has also been obtained within the 16-site numerical calculations in another finite region of the phase diagram. Sec.~\ref{sec:summary} is devoted to discussions and the summary.

\section{Derivation of the effective model}
\label{sec:model}

In this section, we will give a microscopic derivation of the effective pseudospin-$1/2$ Hamiltonian in a comprehensive manner. Though we focus on $4f$ localized moments of Pr$^{3+}$ ions, the form of our nearest-neighbor anisotropic pseudospin-$1/2$ Hamiltonian is most generic for atomic non-Kramers magnetic doublets. We will also present the generic form of the nearest-neighbor Hamiltonian for Kramers doublets of Nd$^{3+}$, Sm$^{3+}$, and Yb$^{3+}$.

\subsection{Atomic Hamiltonian for Pr$^{3+}$}
\label{sec:model:local}

\subsubsection{Coulomb repulsion}
\label{sec:model:Coulomb}

\begin{figure}[tbh]
\begin{center}\leavevmode
\includegraphics[width=\columnwidth]{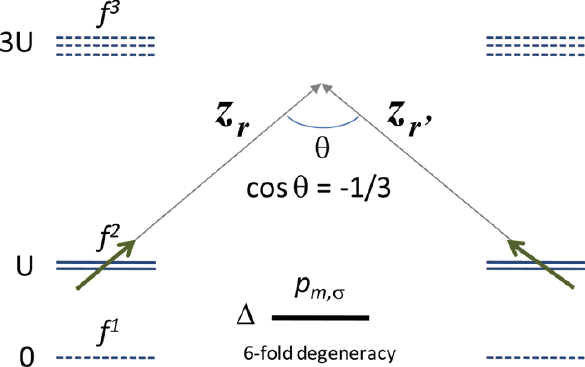}
\end{center}
\caption{(Color online) Local level scheme for $f$ and $p$ electrons, and the local quantization axes $\bm{z}_{\bm{r}}$ and $\bm{z}_{\bm{r'}}$.}
\label{fig:atomic}
\end{figure}

The largest energy scale of the problem should be the local Coulomb repulsion among Pr $4f$ electrons. A photoemission spectroscopy on Pr$_2TM_2$O$_7$, which is desirable for its reliable estimate, is not available yet. A typical value obtained from Slater integrals for Pr$^{3+}$ ions is of order of 3-5~eV~\cite{norman:95}. A cost of the Coulomb energy becomes $0$, $U$, and $3U$ for the occupation of one, two, and three $f$ electrons, respectively [Fig.~\ref{fig:atomic}]. For Pr$^{3+}$ ions, the O $2p$ electron level $\Delta$ at the O1 site should be higher than the $f^1$ level and lower than the $f^3$ level [Fig.~\ref{fig:atomic}].  Then, it would be a reasonably good approximation to start from localized $f^2$ states for Pr$^{3+}$ configurations and then to treat the other effects as perturbations.

\subsubsection{$LS$~coupling for $f^2$ configurations}
\label{sec:mode:LS}

We introduce operators $\hat{\bm{J}}$, $\hat{\bm{L}}$, and $\hat{\bm{S}}$ for the total, orbital, and spin angular momenta of $f^2$ electron states of Pr$^{3+}$. Within this $f^2$ manifold, the predominant $LS$ coupling $\lambda_{LS}>0$ in
$\hat{H}_{LS}=\lambda_{LS}\hat{\bm{L}}\cdot\hat{\bm{S}}$
gives the ground-state manifold ${}^3H_4$ with the quantum numbers $J=4$, $L=5$, and $S=1$ for the total, orbital, and spin angular momenta, respectively. 

\subsubsection{Crystalline electric field}
\label{sec:model:CEF}

The ninefold degeneracy of the ground-state manifold $^3H_4$ is partially lifted by the local crystalline electric field (CEF), which has the $D_{3d}$ symmetry about the $\langle111\rangle$ direction toward the O1 site. We define the local quantization axis $\bm{z}_{\bm{r}}$ as this $\langle111\rangle$ direction. Then, the Hamiltonian for the CEF, 
\begin{equation}
  \hat{H}_{\mathrm{CEF}} = \sum_{m_l,m_l'=-3}^3V_{\mathrm{CEF}}^{m_l,m_l'}\sum_{\bm{r}}\sum_{\sigma=\pm}f^\dagger{}_{\bm{r},m_l,\sigma}f_{\bm{r},m_l',\sigma},
  \label{eq:H_cry}
\end{equation}
contains not only orthogonal components with $m_l=m_l'$ but also off-diagonal components with $m_l-m_l'=\pm3$ and $\pm6$, all of which become real if we take $x$ and $y$ axes as $\bm{x}_{\bm{r}}$ and $\bm{y}_{\bm{r}}$ depicted in Figs.~\ref{fig:crystal} (a) and (b). 
Here, $f^{}_{\bm{r},m_l,\sigma}$ and $f^\dagger_{\bm{r},m_l,\sigma}$ denote the annihilation and creation operators of an $f$ electron with the $z$ components $m_l$ and $m_s=\sigma/2$ of the orbital and spin angular momenta, respectively, in the local coordinate frame at the Pr site $\bm{r}$. The formal expressions for $V_{\mathrm{CEF}}^{m_l,m_l'}$ within the point-charge analysis are given in Appendix~\ref{app:CEF}. In the rest of Sec.~\ref{sec:model:local}, we drop the subscript for the site $\bm{r}$ for brevity. 

We perform the first-order degenerate perturbation theory, which replaces Eq.~\eqref{eq:H_cry} with $\hat{P}({}^3H_4)\hat{H}_{\mathrm{CEF}}\hat{P}({}^3H_4)$, where $\hat{P}({}^3H_4)$ is the projection operator onto the ${}^3H_4$ manifold.
First, let us introduce a notation of $|L,M_L;S,M_S\rangle$ for the $f^2$ eigenstate corresponding to the orbital and spin quantum numbers $(L, M_L)$ and $(S,M_S)$ in the local coordinate frame.
It is straightforward to express the eigenstates $\{|M_J\rangle\}_{M_J=-J,\cdots,J}$ of $\hat{J}^z$  in terms of $|5,M_L;1,M_S\rangle$ and then in terms of $f$-electron operators,
\begin{eqnarray}
  |M_J\rangle
  &=&\sum_{M_L,M_S} C_{M_J,M_L,M_S}|5,M_L;1,M_S\rangle
  \nonumber\\
  &=&\sum_{M_L,M_S} \tilde{C}_{M_J,m,m',\sigma,\sigma'}f^\dagger_{m,\sigma}f^\dagger_{m',\sigma'}|0\rangle,
  \label{eq:J^z=M_J}
\end{eqnarray}
as explicitly written in Appendix~\ref{app:f}.

Finally, we obtain the following representation of Eq.~\eqref{eq:H_cry} in terms of Eq.~\eqref{eq:J^z=M_J}
\begin{widetext}
\begin{eqnarray}
  \langle M_J|\hat{H}_{\mathrm{CEF}}|M_J'\rangle
  &=&\sum_{m_l,m_l',m_l''=-3}^3V_{\mathrm{CEF}}^{m_l,m_l'}
  \sum_{\sigma,\sigma'=\pm}\left[
    \tilde{C}_{M_J,m_l,m_l'',\sigma,\sigma'}\tilde{C}_{M_J',m_l',m_l'',\sigma,\sigma'}
    -\tilde{C}_{M_J,m_l'',m_l,\sigma,\sigma'}\tilde{C}_{M_J',m_l',m_l'',\sigma',\sigma}
    \right.\nonumber\\
    &&\left.\ \ \ \ \ \ \ \ \ \ \ \ \ \ \ \ \ \ \ \ \ \ \ \ \ \ \ \ \ \ \ \ \ \ \ 
    -\tilde{C}_{M_J,m_l,m_l'',\sigma,\sigma'}\tilde{C}_{M_J',m_l'',m_l',\sigma',\sigma}
    +\tilde{C}_{M_J,m_l'',m_l,\sigma,\sigma'}\tilde{C}_{M_J',m_l'',m_l',\sigma,\sigma'}
    \right].\ \ \ 
\end{eqnarray}
\end{widetext}

The CEF favors $M_J=\pm 4$ configurations that are linearly coupled to $M_J=\pm1$ and $\mp2$ because of the $D_{3d}$ CEF. This leads to the atomic non-Kramers magnetic ground-state doublet,
\begin{equation}
  |\sigma\rangle_D=\alpha|4\sigma\rangle+\beta\sigma|\sigma\rangle-\gamma|-2\sigma\rangle,
\label{eq:local}
\end{equation}
with small real coefficients $\beta$ and $\gamma$ as well as $\alpha=\sqrt{1-\beta^2-\gamma^2}$. 
For Pr$_2$Ir$_2$O$_7$, the first CEF excited state is a singlet located at 168~K and the second is a doublet at 648~K~\cite{machida:phd}. They are located at 210~K and 430~K for Pr$_2$Sn$_2$O$_7$~\cite{zhou:08}. These energy scales are two orders of magnitude larger than that of our interest, $2J_{\mathrm{n.n.}}\sim1.4$~K. Hence it is safe to neglect these CEF excitations for our purpose. 
Then, it is convenient to introduce the Pauli matrix vector $\hat{\bm{\sigma}}_{\bm{r}}$ for the pseudospin-$1/2$ representing the local doublet at each site $\bm{r}$, so that Eq.~\eqref{eq:local} is the eigenstate of $\hat{\sigma}^z_{\bm{r}}=\hat{\bm{\sigma}}_{\bm{r}}\cdot\bm{z}_{\bm{r}}$ with the eigenvalue $\sigma$.

Note that in the case of Tb$^{3+}$, the first CEF excited state is a doublet at a rather low energy $\sim18.7$~K~\cite{gingras:00}, and the effects of the first excited doublet cannot be ignored~\cite{molavian:07} when a similar analysis is performed. Nevertheless, since this CEF excitation in Tb$^{3+}$ is an order of magnitude larger than $J_{\mathrm{eff}}$, it could be integrated out~\cite{molavian:07}. Then, the model reduces to a similar form of the effective pseudospin-$1/2$ Hamiltonian which has been derived in Ref.~\onlinecite{onoda:09} and will also be discussed below, though the explicit form has not been presented as far as we know.

\subsection{Dipole and quadrupole moments}
\label{sec:model:moment}

Only the $z$ component $\hat{\sigma}^z_{\bm{r}}=\hat{\bm{\sigma}}_{\bm{r}}\cdot\bm{z}_{\bm{r}}$ of the pseudospin contributes to the {\em magnetic dipole moment} represented as either ``in'' or ``out'', while the transverse components $\hat{\sigma}^x=\hat{\bm{\sigma}}_{\bm{r}}\cdot\bm{x}_{\bm{r}}$ and $\hat{\sigma}^y_{\bm{r}}=\hat{\bm{\sigma}}_{\bm{r}}\cdot\bm{y}_{\bm{r}}$ correspond to the {\em atomic quadrupole moment}, i.e., the orbital. 
This can be easily shown by directly calculating the Pr $4f$ magnetic dipole and quadrupole moments in terms of the pseudospin. We first take the projection of the total angular momentum $\hat{\bm{J}}=(\hat{J}^x,\hat{J}^y,\hat{J}^z)$ to the subspace of the local non-Kramers magnetic ground-state doublet described by Eq.~(\ref{eq:local}). It yields
\begin{eqnarray}
_D\langle\sigma|\hat{J}^z|\sigma'\rangle_D&=&(4\alpha^2+\beta^2-2\gamma^2)\sigma\delta_{\sigma,\sigma'},
\label{eq:Jz}\\
_D\langle\sigma|\hat{J}^\pm|\sigma'\rangle_D&=&0,
\label{eq:J+-}
\end{eqnarray}
with $\hat{J}^\pm=(\hat{J}^x\pm i\hat{J}^y)$. 
With the Land\'{e} factor $g_J=4/5$ and the Bohr magneton $\mu_B$, the atomic magnetic dipole moment is given by
\begin{equation}
  \hat{\bm{m}}_{\bm{r}}=g_J\mu_B(4\alpha^2+\beta^2-2\gamma^2)\hat{\sigma}^z_{\bm{r}}\bm{z}_{\bm{r}}.
  \label{eq:m}
\end{equation}
Note that $\hat{\sigma}_{\bm{r}}^\pm$ cannot linearly couple to neutron spins without resorting to higher CEF levels. 
On the other hand, the quadrupole moments are given by 
\begin{equation}
_D\langle\sigma|\{\hat{J^z},\hat{J}^\pm\}|\sigma'\rangle_D=-36\beta\gamma\delta_{\sigma,-\sigma'}.
\label{eq:JxJ+-}
\end{equation}

This is a general consequence of the so-called non-Kramers magnetic doublet and not restricted to the Pr$^{3+}$ ion. Namely, when the atomic ground states of the non-Kramers ions having an even number of $f$ electrons and thus an integer total angular momentum $J$ are described by a magnetic doublet, only $\hat{\sigma}^z_{\bm{r}}$ contributes to the {\em magnetic dipole moment}, while $\hat{\sigma}^{x,y}_{\bm{r}}$ corresponds to the {\em atomic quadrupole moment}. 
This sharply contrasts to the following two cases:
(i) In the case of Kramers doublets, all three components of $\hat{\bm{\sigma}}_{\bm{r}}$ may contribute to the magnetic dipole moment while their coefficients can be anisotropic.
(ii) In the case of non-Kramers non-magnetic doublets, $\hat{\bm{\sigma}}_{\bm{r}}$ corresponds to a quadrupole moment or even a higher-order multipole moment which is a time-reversal invariant.

\subsection{Superexchange interaction}
\label{sec:model:superexchange}

Now we derive the superexchange Hamiltonian through the fourth-order strong-coupling perturbation theory. Keeping in mind the local level scheme of Pr $4f$ electrons and O $2p$ electrons at O1 sites, which has been explained in Sec.~\ref{sec:model:local}, we consider nonlocal effects introduced by the electron transfer between the Pr $4f$ orbital and the O $2p$ orbital. 

\subsubsection{Local coordinate frames}
\label{sec:model:frames}

In order to symmetrize the final effective Hamiltonian, it is convenient to choose a set of local coordinate frames so that it is invariant under $180^\circ$ rotations of the whole system about three axes that include an O1 site and are parallel to the global $X$, $Y$, or $Z$ axes, which belong to the space group $F_{d\bar{3}m}$ of the present pyrochlore system. We can start from the local coordinate frame previously defined in Sec.~\ref{sec:model:CEF} and in Fig.~\ref{fig:crystal} for a certain site and generate the other three local frames by applying the above three rotations. 

For instance, we can adopt
\begin{subequations}
  \begin{eqnarray}
    \bm{x}_0&=&\frac{1}{\sqrt{6}}\left(1,1,-2\right),
    \nonumber\\
    \bm{y}_0&=&\frac{1}{\sqrt{2}}\left(-1,1,0\right),
    \nonumber\\
    \bm{z}_0&=&\frac{1}{\sqrt{3}}(1,1,1),
    \label{eq:xyz0}
  \end{eqnarray}
  for the Pr sites at $\bm{R}+\bm{a}_0$ with $\bm{a}_0=-\frac{a}{8}(1,1,1)$,
  \begin{eqnarray}
    \bm{x}_1&=&\frac{1}{\sqrt{6}}\left(1,-1,2\right),
    \nonumber\\
    \bm{y}_1&=&\frac{1}{\sqrt{2}}\left(-1,-1,0\right),
    \nonumber\\
    \bm{z}_1&=&\frac{1}{\sqrt{3}}(1,-1,-1),
    \label{eq:xyz1}
  \end{eqnarray}
  for the Pr sites at $\bm{R}+\bm{a}_1$ with $\bm{a}_1=\frac{a}{8}(-1,1,1)$,
  \begin{eqnarray}
    \bm{x}_2&=&\frac{1}{\sqrt{6}}\left(-1,1,2\right),
    \nonumber\\
    \bm{y}_2&=&\frac{1}{\sqrt{2}}\left(1,1,0\right),
    \nonumber\\
    \bm{z}_2&=&\frac{1}{\sqrt{3}}(-1,1,-1),
    \label{eq:xyz2}
  \end{eqnarray}
  for the Pr sites at $\bm{R}+\bm{a}_2$ with $\bm{a}_2=\frac{a}{8}(1,-1,1)$, and
  \begin{eqnarray}
    \bm{x}_3&=&\frac{1}{\sqrt{6}}\left(-1,-1,-2\right),
    \nonumber\\
    \bm{y}_3&=&\frac{1}{\sqrt{2}}\left(1,-1,0\right),
    \nonumber\\
    \bm{z}_3&=&\frac{1}{\sqrt{3}}(-1,-1,1),
    \label{eq:xyz3}
  \end{eqnarray}
  for the Pr sites at $\bm{R}+\bm{a}_3$ with $\bm{a}_3=\frac{a}{8}(1,1,-1)$,
  \label{eq:xyz}
\end{subequations}
where $\bm{R}$ represents a fcc lattice vector $\bm{R}=\sum_{i=1,2,3}n_i\bm{R}_i$ spanned by $\bm{R}_1=(0,a/2,a/2)$, $\bm{R}_2=(a/2,0,a/2)$, and $\bm{R}_3=(a/2,a/2,0)$ with integers $(n_1,n_2,n_3)$, and $a$ is the lattice constant, i.e., the side length of the unit cube. In particular, all the local $z$ directions attached to the Pr sites belonging to the tetrahedron centered at the O1 site $\bm{R}$ point inwards, and they satisfy the relation
\begin{equation}
  \sum_{i=0}^3(\bm{x}_i,\bm{y}_i,\bm{z}_i)=(\bm{0},\bm{0},\bm{0}).
  \label{eq:sum_xyz}
\end{equation}
Actually, other sets of local coordinate frames which are obtained by threefold and sixfold rotations about $[111]$ yield exactly the same expression for the effective Hamiltonian for Kramers and non-Kramers cases, respectively. 

These local coordinate frames are related to the following rotations of the global coordinate frame,
\begin{subequations}
  \begin{eqnarray}
  R^{(r)}(\varphi_i,\vartheta_i)
  &=&\left(
    {}^t\bm{x}_i,
    {}^t\bm{y}_i,
    {}^t\bm{z}_i
  \right),
  \\
    \varphi_0=\pi/4,&\ \ \ &
    \vartheta_0=\arccos\left(1/\sqrt{3}\right),
    \label{eq:angles0}\\
    \varphi_1=3\pi/4,&\ \ \ &
    \vartheta_1=-\pi+\arccos\left(1/\sqrt{3}\right),
    \label{eq:angles1}\\
    \varphi_2=-\pi/4,&\ \ \ &
    \vartheta_2=-\pi+\arccos\left(1/\sqrt{3}\right),
    \label{eq:angles2}\\
    \varphi_3=-3\pi/4,&\ \ \ &
    \vartheta_3=\arccos\left(1/\sqrt{3}\right).
    \label{eq:angles3}
  \end{eqnarray}
\label{eq:R^r}
\end{subequations}
Note that the coordinate frame for the spins is always attached to that for the orbital space in each case. The rotation of $\bm{j}=\bm{l}+\bm{s}$ with the orbital $\bm{l}$ and the spin $\bm{s}$ of a single electron takes the form
\begin{equation}
  \hat{R}_{\bm{r}}=\exp\left[-i\varphi_i\hat{j}^z\right]\exp\left[-i\vartheta_i\hat{j}^y\right].
  \label{eq:R}
\end{equation}

\subsubsection{$f$-$p$ hybridization}
\label{sec:model:hybridization}

\begin{figure}[tbh]
\begin{center}\leavevmode
\includegraphics[width=\columnwidth]{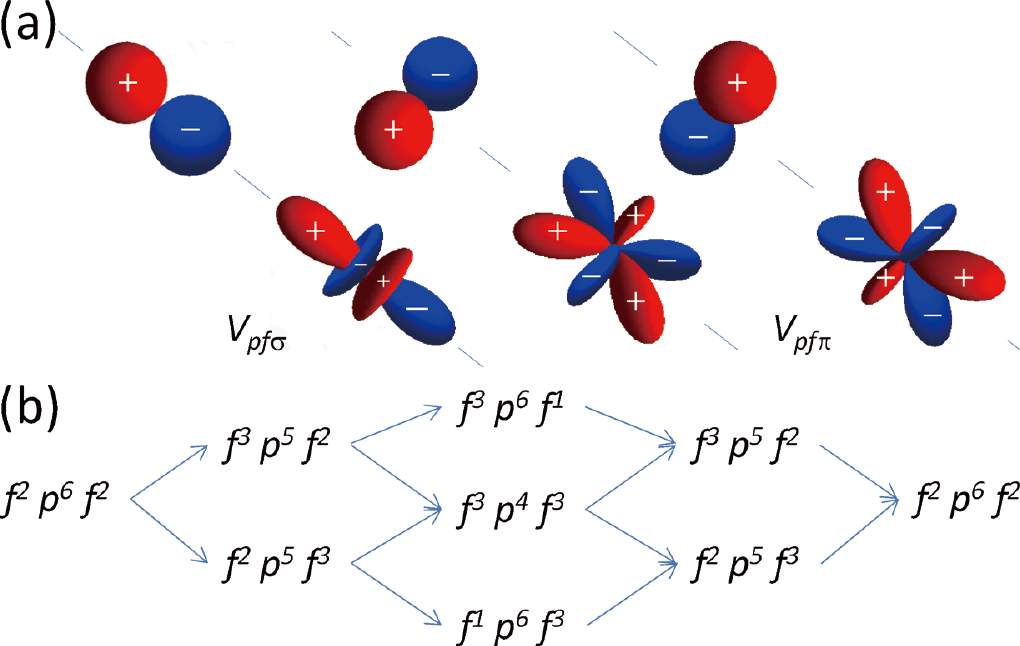}
\end{center}
\caption{(Color online) (a) Two $f$-$p$ transfer integrals; $V_{pf\sigma}$ between $p_z$ and $f_{(5z^2-3r^2)z}$ orbitals, and $V_{pf\pi}$ between $p_x$/$p_y$ and $f_{x(5z^2-r^2)}$/$f_{y(5z^2-r^2)}$. (b) $f$-$p$ virtual electron hopping processes. $n$ ($n'$) and $\ell$ in the state $f^np^\ell f^{n'}$ represent the number of $f$ electrons at the Pr site $\bm{r}$ ($\bm{r}'$) and that of $p$ electrons at the O1 site.}
\label{fig:perturbation}
\end{figure}

The $4f$ electrons occupying the atomic ground-state doublet, Eq.~\eqref{eq:local}, or $4f$ holes can hop to the O $2p$ levels at the neighboring O1 site. Because of the symmetry, the $f$-$p$ electron transfer along the local $\bm{z}$ axis is allowed only for the $pf\sigma$ bonding ($m_l=0$) between $f_{z(5z^2-3r^2)}$ and $p_z$ orbitals and the $pf\pi$ bondings ($m_l=\pm1$) between $f_{x(5z^2-r^2)}$ and $p_x$ orbitals and between $f_{y(5z^2-r^2)}$ and $p_y$ orbitals in the local  coordinate frame defined in Eqs.~\eqref{eq:xyz} [Fig.~\ref{fig:perturbation} (a)]. Their amplitudes are given by two Slater-Koster parameters~\cite{sharma:79} $V_{pf\sigma}$ and $V_{pf\pi}$, respectively. Then, the Hamiltonian for the $f$-$p$ hybridization reads
\begin{widetext}
\begin{equation}
  \hat{H}_{\mathrm{t}}=\sum_{\bm{R}\in fcc}\sum_{\tau=\pm}\,\sum_{m_l,m_l'=0,\pm1}V_{m_l}
  \sum _{\sigma,\sigma' =\pm}
  \hat{f}^{\dagger }{}_{\bm{R}+\bm{a}_i,m_l,\sigma}
  \left(R^\dagger_{\bm{R}+\bm{a}_i}\right)_{m_l,m_l';\sigma,\sigma'}
  \hat{p}_{\bm{R}+(1+\tau)\bm{a}_i,m_l',\sigma'}
  +h.c.
  \label{eq:H_t}
\end{equation}
\end{widetext}
with $V_{\pm1}=V_{pf\pi}$, $V_0=V_{pf\sigma}$, and an index $\tau=\pm$ for the two fcc sublattices of the diamond lattice, where $\hat{p}_{\bm{R}+(1+\tau)\bm{a}_i,m_l,\sigma}$ represents the annihilation operator of a $2p$ electron at the O1 site $\bm{R}+(1+\tau)\bm{a}_i$ with the orbital and spin quantum numbers $m_l$ and $m_s=\sigma/2$, respectively, in the global coordinate frame. Here, $R^\dagger_{\bm{R}+\bm{a}_i}$ transforms the representation from the global frame for $\hat{p}_{\bm{R}+(1+\tau)\bm{a}_i,m_l',\sigma'}$ to the local frame for $\hat{f}^{\dagger }{}_{\bm{R}+\bm{a}_i,m_l,\sigma}$.

\subsubsection{Strong-coupling perturbation theory}
\label{sec:mode:perturbation}

Now we are ready to perform the strong-coupling perturbation expansion in $V_{pf\pi}$ and $V_{pf\sigma}$. Hybridization between these Pr $4f$ electrons and O $2p$ electrons at the O1 site, which is located at the center of the tetrahedron, couples $f^2$ states having the local energy $U$ with $f^1$ and $f^3$ states having the local energy levels 0 and $3U$, respectively [Fig.~\ref{fig:atomic}]. Here, the $LS$ coupling has been ignored in comparison with $U$ for simplicity. Creating a virtual $p$ hole decreases the total energy by $\Delta$, which is the $p$ electron level measured from the $f^1$ level. 

First, the second-order perturbation in $V_{pf\sigma}$ and $V_{pf\pi}$ produces only local terms. They only modify the CEF from the result of the point-charge analysis with renormalized parameters for the effective ionic charges and radii. 
Nontrivial effects appear in the fourth order in $V_{pf\sigma}$ and $V_{pf\pi}$. Taking into account the virtual processes shown in Fig.~\ref{fig:perturbation} (b), the fourth-order perturbed Hamiltonian in $V_{pf\sigma}$ and $V_{pf\pi}$ is obtained as
\begin{widetext}
\begin{eqnarray}
  \hat{H}_{ff}&=&\frac{2}{(2U-\Delta)^2}\sum_{\langle\bm{r},\bm{r}'\rangle}^{\mathrm{n.n.}}\sum_{m_1,m_2}\sum_{m_1',m_2'}\sum_{\sigma_1,\sigma_2}\sum_{\sigma_1',\sigma_2'}
    V_{m_1}V_{m_1'}V_{m_2}V_{m_2'}
    \hat{f}^\dagger{}_{\bm{r},m_1,\sigma_1}\hat{f}_{\bm{r},m_2,\sigma_2}
    \hat{f}^\dagger{}_{\bm{r}',m_1',\sigma_1'}\hat{f}_{\bm{r}',m_2',\sigma_2'}
  \nonumber\\
  &&\times\biggl[-\frac{1}{2U-\Delta}\delta_{m_1,m_2}\delta_{m_1',m_2'}\delta_{\sigma_1,\sigma_2}\delta_{\sigma_1',\sigma_2'}
    +(\frac{1}{2U-\Delta}+\frac{1}{U})\left(R^\dagger_{\bm{r}}R_{\bm{r}'}\right)_{m_1,m_2';\sigma_1,\sigma_2'}\left(R^\dagger_{\bm{r}'}R_{\bm{r}}\right)_{m_1',m_2;\sigma_1',\sigma_2}
\biggr].
  \label{eq:H_sex}
\end{eqnarray}
\end{widetext}

\subsection{Effective pseudospin-$1/2$ model}
\label{sec:model:hamiltonian}

\begin{figure}
\begin{center}
\includegraphics[width=\columnwidth]{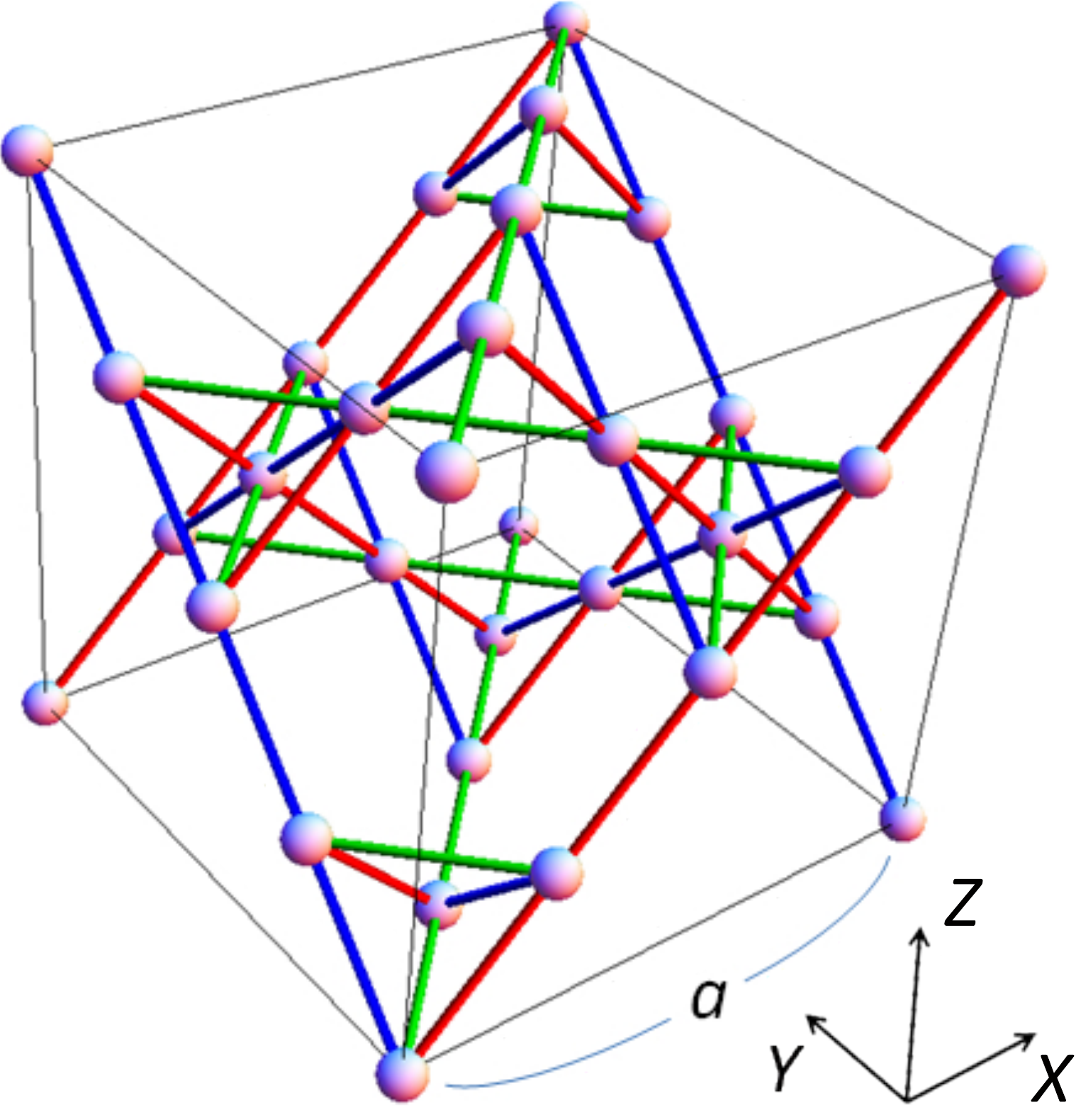}
\end{center}
\caption{(Color online) The pyrochlore lattice structure. The phase $\phi_{\bm{r},\bm{r}'}$ appearing in Eq.~(\ref{eq:H_eff}) takes $-2\pi/3$, $2\pi/3$, and $0$ on the blue, red, green bonds, respectively, in our choice of the local coordinate frames [Eqs.~\ref{eq:xyz}].}
\label{fig:pyrochlore}
\end{figure}

Next we project the superexchange Hamiltonian Eq.~\eqref{eq:H_sex} onto the subspace of doublets given by Eq.~(\ref{eq:local}). For this purpose, we have only to calculate for a site $\bm{r}$ the matrix elements of the operators $\hat{f}^\dagger{}_{\bm{r},m_l,\sigma}\hat{f}_{\bm{r},m_l',\sigma'}$ with $m_l,m_l'=0,\pm1$ and $\sigma,\sigma'=\pm1$, in terms of $|M_J\rangle$ that is explicitly represented with $f$-electron operators in Appendix~\ref{app:f}, and then in terms of the atomic doublet $|\sigma\rangle_D$, Eq.~\eqref{eq:local}. 
Then, we finally obtain the effective quantum pseudospin-$1/2$ Hamiltonian;
\begin{eqnarray}
  \hat{H}_{\mathrm{eff}}&=&J_{\mathrm{n.n.}}\sum_{\langle\bm{r},\bm{r}'\rangle}^{\mathrm{n.n.}}\left[\hat{\sigma}_{\bm{r}}^z\hat{\sigma}_{\bm{r}'}^z+2\delta\left(\hat{\sigma}_{\bm{r}}^+\hat{\sigma}_{\bm{r}'}^-+\hat{\sigma}_{\bm{r}}^-\hat{\sigma}_{\bm{r}'}^+\right)
    \right.
    \nonumber\\
    &&\left.+2q\left(e^{2i\phi_{\bm{r},\bm{r}'}}\hat{\sigma}_{\bm{r}}^+\hat{\sigma}_{\bm{r}'}^++h.c.\right)\right],
  \label{eq:H_eff}
\end{eqnarray}
with $\hat{\sigma}^\pm_{\bm{r}}\equiv(\hat{\sigma}^x_{\bm{r}}\pm i\hat{\sigma}^y_{\bm{r}})/2$,
where $\hat{\bm{\sigma}}_{\bm{r}}$ represents a vector of the Pauli matrices for the pseudospin at a site $\bm{r}$. The phase~\cite{phase} $\phi_{\bm{r},\bm{r}'}$ takes $-2\pi/3$, $2\pi/3$, and $0$ for the bonds shown in blue, red, and green colors in Fig.~\ref{fig:pyrochlore}, in the local coordinate frames defined in Eq.~\eqref{eq:xyz}. This phase cannot be fully gauged away, because of the noncollinearity of the $\langle111\rangle$ magnetic moment directions and the threefold rotational invariance of $(\bm{r}, \bm{\sigma}_{\bm{r}})$ about the [111] axes. 
 Equation~\eqref{eq:H_eff} gives the most generic nearest-neighbor pseudospin-$1/2$ Hamiltonian for non-Kramers magnetic doublets of rare-earth ions such as Pr$^{3+}$ and Tb$^{3+}$ that is allowed by the symmetry of the pyrochlore system. Note that the bilinear coupling terms of $\hat{\sigma}^z_{\bm{r}}$ and $\hat{\sigma}^\pm_{\bm{r}'}$ are prohibited by the non-Kramers nature of the moment; namely $\hat{\sigma}^z_{\bm{r}}$ changes the sign under the time-reversal operation, while $\hat{\sigma}^{x,y}_{\bm{r}}$ does not.

\begin{figure}[tbh]
\begin{center}
\includegraphics[width=\columnwidth]{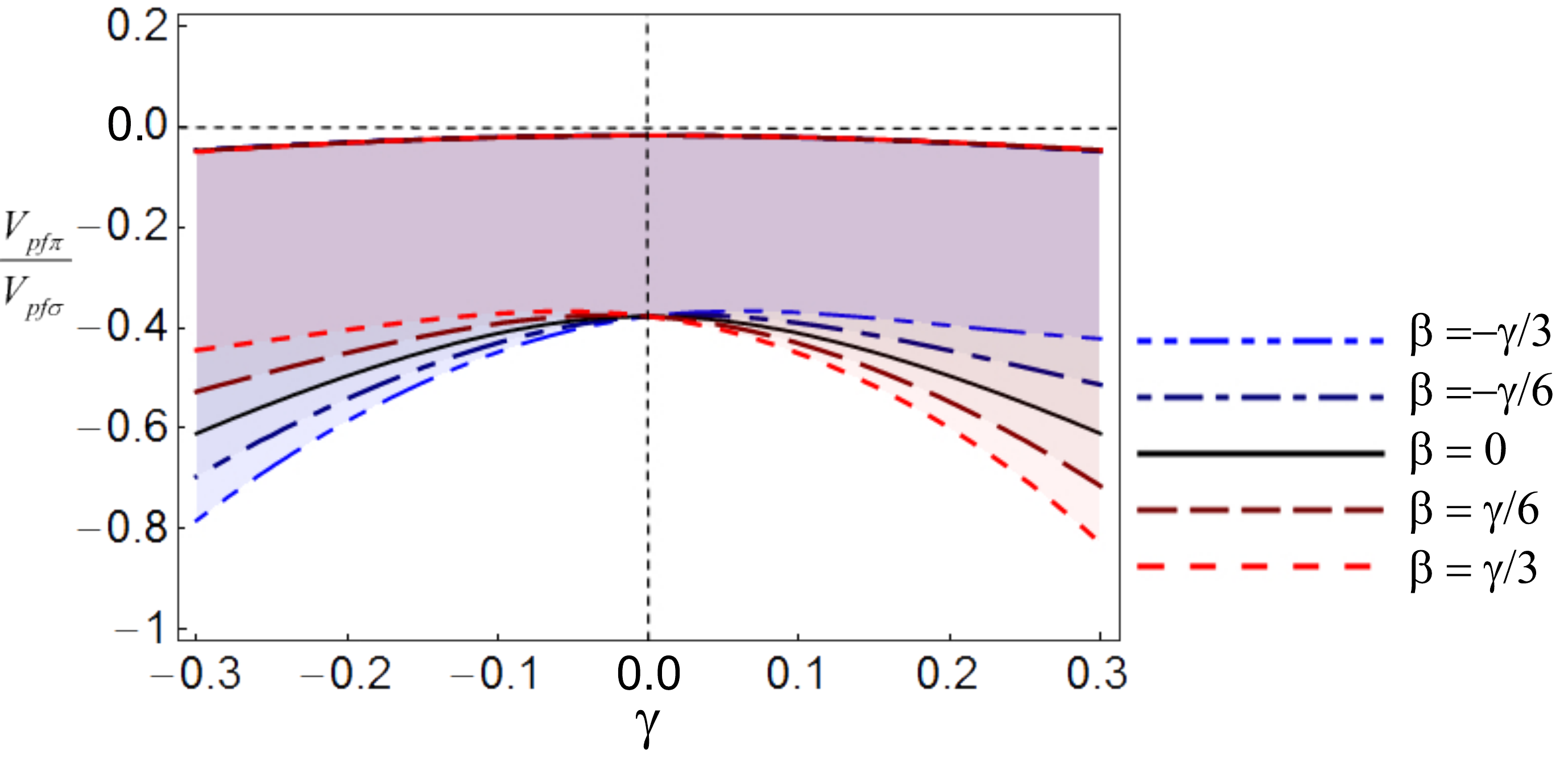}
\end{center}
\caption{(Color online) A phase diagram for the sign of the dimensionless Ising coupling $\tilde{J}$, defined through Eq.~\eqref{eq:J}, as functions of $\gamma$ and $V_{pf\pi}/V_{pf\sigma}$ for $\beta=\gamma/3$, $\gamma/6$, $0$, $-\gamma/6$, and $-\gamma/3$. In each case, $\tilde{J}$ is positive in the shaded region, and negative otherwise.}
\label{fig:J}
\end{figure}

The dependence of the Ising coupling constant on $U$, $\Delta$, $V_{pf\sigma}$ and $V_{pf\pi}$  takes the form,
\begin{equation}
  J_{\mathrm{n.n.}}=\frac{V_{pf\sigma}^4}{(2U-\Delta)^2}\left(\frac{1}{U}+\frac{1}{2U-\Delta}\right)\tilde{J}(\beta,\gamma,V_{pf\pi}/V_{pf\sigma}),
  \label{eq:J}
\end{equation}
where $\tilde{J}$ contains the dependence on the remaining dimensionless variables $(\beta,\gamma,V_{pf\pi}/V_{pf\sigma})$. We show the sign of $\tilde{J}$ as functions of $\gamma$ and $V_{pf\pi}/V_{pf\sigma}$ for several choices of $\beta/\gamma$ in Fig.~\ref{fig:J}.
In particular, for $-0.37\lesssim V_{pf\pi}/V_{pf\sigma}\lesssim-0.02$ which includes a realistic case of $V_{pf\pi}/V_{pf\sigma}\sim-0.3$, $\tilde{J}$ is found to be positive. Since the prefactor in Eq.~\eqref{eq:J} is positive, the Ising coupling $J_{\mathrm{n.n.}}$ is also positive, namely, antiferroic for pseudospins, in this case. Taking account of the tilting of the two neighboring local $z$ axes by $\theta=\arccos(-1/3)$, this indicates a ferromagnetic coupling between the physical $\langle111\rangle$ magnetic moments and provides a source of the ice rule.

\begin{figure}[tbh]
\begin{center}
\includegraphics[width=\columnwidth]{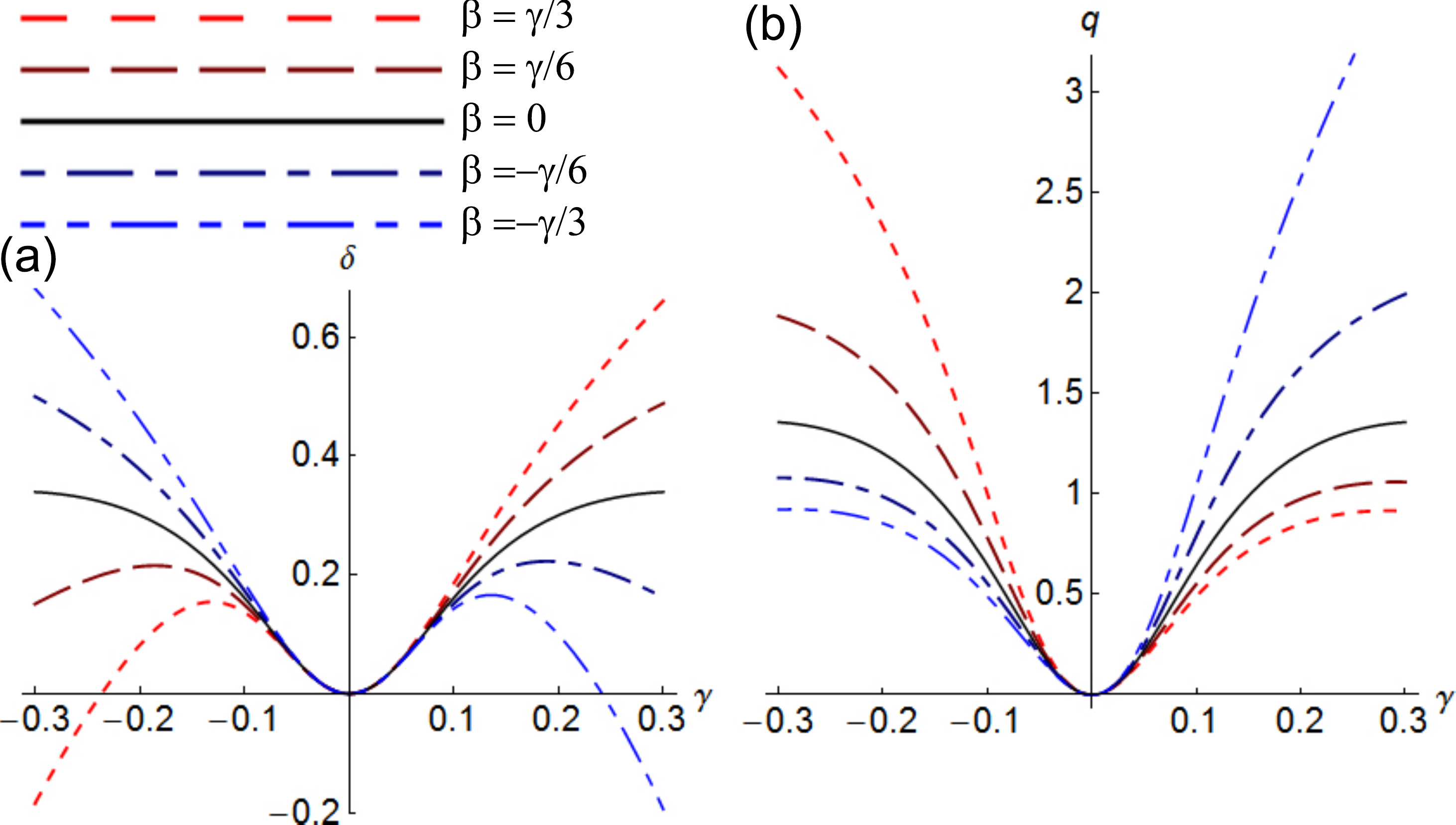}
\end{center}
\caption{(Color online) The coupling constants (a) $\delta$ and (b) $q$ as functions of $\gamma$ for several choices of $\beta/\gamma=1/3$, $1/6$, $0$, $-1/6$, and $-1/3$. We have adopted $U/V_{pf\sigma}=5$, $\Delta/V_{pf\sigma}=4$, and $V_{pf\pi}/V_{pf\sigma}=-0.3$.}
\label{fig:coupling}
\end{figure}

The $D_{3d}$ CEF produces two quantum-mechanical interactions in the case of non-Kramers ions; the pseudospin-exchange and pseudospin-nonconserving terms. The ratios $\delta$ and $q$ of their coupling constants to the Ising one are insensitive to $U/V_{pf\sigma}$ and $\Delta/V_{pf\sigma}$ but strongly depends on $\beta$ and $\gamma$. Figures~\ref{fig:coupling} (a) and (b) show $\delta$ and $q$, respectively, as functions of $\gamma$ that characterizes the $D_{3d}$ CEF for a typical choice of parameters, $U/V_{pf\sigma}=5$, $\Delta/V_{pf\sigma}=4$, and $V_{pf\pi}/V_{pf\sigma}=-0.3$, in the cases of $\beta/\gamma=0, \pm1/3, \pm1/6$. Henceforth, we adopt $\beta=7.5\%$ and $\beta/\gamma=1/3$, following the point-charge analysis of the inelastic powder neutron-scattering data on Pr$_2$Ir$_2$O$_7$~[Ref.~\onlinecite{machida:phd}]. Actually, these estimates of $\beta$ and $\gamma$ lead to the local moment amplitude,
\begin{equation}
  M_0=g_J\mu_B(4\alpha^2+\beta^2-2\gamma^2)\approx2.9\mu_B,
  \label{eq:M_0}
\end{equation}
according to Eq.~\eqref{eq:m}, which reasonably agrees with the experimental observation on Pr$_2$Ir$_2$O$_7$~\cite{nakatsuji:06} and Pr$_2$Zr$_2$O$_7$~\cite{matsuhira:09}. Then, we obtain $\delta\sim0.51$ and $q\sim0.89$, indicating the appreciable quantum nature. In general, however, the values of $\delta$ and $q$ may vary depending on a transition-metal ion and crystal parameters.
We note that a finite $q$ was not taken into account in the literature seriously.

It is instructive to rewrite the $q$ term as
\begin{eqnarray}
q
\left((\hat{\vec{\sigma}}_{\bm{r}}\cdot\vec{n}_{\bm{r},\bm{r}'})(\hat{\vec{\sigma}}_{\bm{r}'}\cdot\vec{n}_{\bm{r},\bm{r}'}) 
-(\hat{\vec{\sigma}}_{\bm{r}}\cdot\vec{n}'_{\bm{r},\bm{r}'})(\hat{\vec{\sigma}}_{\bm{r}}\cdot\vec{n}'_{\bm{r},\bm{r}'})\right).
\label{eq:q2}
\end{eqnarray}
where we have introduced a two-dimensional vector composed of the planar components of the pseudospin, $\hat{\vec{\sigma}}_{\bm{r}}=(\hat{\sigma}^x_{\bm{r}},\hat{\sigma}^y_{\bm{r}})$, and two orthonormal vectors 
\begin{subequations}
\begin{eqnarray}
\vec{n}_{\bm{r},\bm{r}'}&=&(\cos\phi_{\bm{r},\bm{r}'},-\sin\phi_{\bm{r},\bm{r}'}),
\label{eq:n}\\
\vec{n}'_{\bm{r},\bm{r}'}&=&(\sin\phi_{\bm{r},\bm{r}'},\cos\phi_{\bm{r},\bm{r}'}). 
\label{eq:n'}
\end{eqnarray}
\end{subequations}
Then, it is clear that the sign of $q$ can be absorbed by rotating all the pseudospins $\bm{\sigma}_{\bm{r}}$ about the local $\bm{z}_{\bm{r}}$ axes by $\pi/2$. Furthermore, in the particular case $\delta=|q|$ or $-|q|$, the planar ($\hat{\vec{\sigma}}_{\bm{r}}$) part, namely, the sum of the $\delta$ and $q$ terms, of the Hamiltonian Eq.~\eqref{eq:H_eff} is reduced to the antiferroic or ferroic pseudospin $120^\circ$ Hamiltonian~\cite{chern:10}.

For Kramers ions such as Nd$^{3+}$, Er$^{3+}$, and Yb$^{3+}$, there appears another coupling constant~\cite{comm,onoda:11} for an additional interaction term
\begin{eqnarray}
  \hat{H}_{\mathrm{K}}=K\sum_{\langle\bm{r},\bm{r}'\rangle}^{\mathrm{n.n.}}\left[
  \hat{\sigma}^z_{\bm{r}}\left(\hat{\vec{\sigma}}_{\bm{r}'}\cdot\vec{n}_{\bm{r},\bm{r}'}\right)
+\left(\hat{\vec{\sigma}}_{\bm{r}}\cdot\vec{n}_{\bm{r},\bm{r}'}\right) \hat{\sigma}^z_{\bm{r}'}\right],
\label{eq:H_K}
\end{eqnarray}
whose form has been obtained so that it satisfies the threefold rotational symmetry about $\langle111\rangle$ axes, i.e., $\bm{z}_{\bm{r}}$, the mirror symmetry about the planes spanned by $\bm{z}_{\bm{r}}$ and $\bm{z}_{\bm{r}'}$ for all the pairs of nearest-neighbor sites $\bm{r}$ and $\bm{r}'$, and the twofold rotational symmetry about $\bm{X}$, $\bm{Y}$, and $\bm{Z}$ axes.
This reflects the fact that all the components of $\hat{\bm{\sigma}}_{\bm{r}}$ change the sign under the time-reversal, $\hat{\bm{\sigma}}_{\bm{r}}\to-\hat{\bm{\sigma}}_{\bm{r}}$, and hence $\hat{H}_{\mathrm{K}}$ respects the time-reversal symmetry for Kramers ions. Equation~\eqref{eq:H_eff} and the additional term Eq.~\eqref{eq:H_K} appearing only for Kramers doublets define the most generic form of the nearest-neighbor bilinear interacting pseudospin-$1/2$ Hamiltonian that is allowed by the symmetry of the magnetic pyrochlore system. Throughout this paper, we restrict ourselves to the case of non-Kramers ions, for which $K$ vanishes and $\hat{H}_{\mathrm{K}}$ does not appear.

\section{Classical mean-field theory}
\label{sec:MFT}

We start with a classical mean-field analysis of our effective Hamiltonian Eq.~\eqref{eq:H_eff} along the strategy of Ref.~\onlinecite{reimers:91}. We look for the instability with decreasing temperature, and then consider candidates to the mean-field ground state in the space of the two coupling constants $\delta$ and $q$. We restrict ourselves to the nearest-neighbor model, though it is known that longer-range interactions can lift the degeneracy at least partially~\cite{reimers:91}.
Since $\hat{\sigma}^z_{\bm{r}}$ and $\hat{\sigma}^{\pm}_{\bm{r}}$ are decoupled in the Hamiltonian in the classical level, we proceed by requiring that either of $\langle\hat{\sigma}^z_{\bm{r}}\rangle$ and/or $\langle\hat{\sigma}^{x,y}_{\bm{r}}\rangle$ is finite. It reveals three distinct mean-field instabilities in the case of $J_{\mathrm{n.n.}}>0$, which also provide candidate mean-field ground states under the constraint $|\langle\hat{\bm{\sigma}}_{\bm{r}}\rangle|\le1$. Note that the decoupling approximation of the Ising and planar components becomes inaccurate when the SU(2)-symmetric point of Heisenberg antiferromagnet.

\subsection{Ising states $\langle\sigma^z_{\bm{r}}\rangle\ne0$}
\label{sec:MFT:dipole}

Let us introduce a vector,
\begin{equation}
  {}^t\hat{d}_{\bm{R}}= \left(\hat{\sigma}^z_{\bm{R}+\bm{a}_0},\hat{\sigma}^z_{\bm{R}+\bm{a}_1},\hat{\sigma}^z_{\bm{R}+\bm{a}_2},\hat{\sigma}^z_{\bm{R}+\bm{a}_3}\right), 
  \label{eq:d}
\end{equation}
where $\bm{R}$ is a fcc lattice vector and $\{\bm{a}_i\}_{i=0,\cdots,3}$ have been defined in Sec.~\ref{sec:model:frames}.
In the mean-field approximation, magnetic dipolar states characterized by a nonzero $\langle\sigma^z_{\bm{r}}\rangle$ is obtained as the states having the minimum eigenvalue of the following mean-field Hamiltonian,
\begin{eqnarray}
  {\cal H}^z_{\mathrm{MF}}
  &=&N_T\sum_{\bm{q}}\langle\hat{d}_{\bm{q}}^\dagger\rangle h^z_{\bm{q}}\langle\hat{d}_{\bm{q}}\rangle
  \label{eq:H^d_MF}\\
  h^z_{\bm{q}}&=&2J_{\mathrm{n.n.}}\left(\begin{array}{cccc}
  0 & f_{q_y + q_z} & f_{q_z + q_x} & f_{q_x + q_y}\\
  f_{q_y + q_z} & 0 & f_{q_x - q_y} & f_{q_z - q_x}\\
  f_{q_z + q_x} & f_{q_x - q_y} & 0 & f_{q_y - q_z}\\
  f_{q_x + q_y} & f_{q_x - q_z} & f_{q_y - q_z} & 0
  \end{array}\right),
  \label{eq:h^d_MF}
\end{eqnarray}
where $f_q=\cos(qa/4)$, and
\begin{equation}
  \hat{d}_{\bm{q}}\equiv\frac{1}{N_T}\sum_{\bm{R}}\left(\begin{array}{c}
\hat{\sigma}^z_{\bm{R}+\bm{a}_0}e^{-i\bm{q}\cdot(\bm{R}+\bm{a}_0)}
\\
\hat{\sigma}^z_{\bm{R}+\bm{a}_1}e^{-i\bm{q}\cdot(\bm{R}+\bm{a}_1)}
\\
\hat{\sigma}^z_{\bm{R}+\bm{a}_2}e^{-i\bm{q}\cdot(\bm{R}+\bm{a}_2)}
\\
\hat{\sigma}^z_{\bm{R}+\bm{a}_3}e^{-i\bm{q}\cdot(\bm{R}+\bm{a}_3)}
\end{array}\right),
  \label{eq:d_q}
\end{equation}
with $N_T=N/4$ where $N$ is the number of pyrochlore-lattice sites. $h^z_{\bm{q}}$ has the eigenvalues~\cite{reimers:91}, 
\begin{equation}
  \varepsilon^z_{\bm{q}}=-2J_{\mathrm{n.n.}},
  2J_{\mathrm{n.n.}}\left(1\mp\sqrt{1+g_{\bm{q}}}\right),
  \label{eq:epsilon^d}
\end{equation}
where $g_{\bm{q}}=f_{2q_x}f_{2q_y}+f_{2q_y}f_{2q_z}+f_{2q_z}f_{2q_x}$.
For $J_{\mathrm{n.n.}}>0$, the lowest energy of the mean-field solution, which is obtained as $-2J_{\mathrm{n.n.}}$ per tetrahedron for any wavevector $\bm{q}$ all over the Brillouin zone reflecting the macroscopic degeneracy, coincides with the exact ground-state energy of the nearest-neighbor spin ice model. 

\subsection{Planar states $\langle\sigma^\pm_{\bm{r}}\rangle\ne0$}
\label{sec:MFT:quadrupole}

Introducing another vector,
\begin{equation}
  {}^t\hat{Q}_{\bm{R}}= (\hat{\vec{\sigma}}_{\bm{R}+\bm{a}_0},\hat{\vec{\sigma}}_{\bm{R}+\bm{a}_1},\hat{\vec{\sigma}}_{\bm{R}+\bm{a}_2},\hat{\vec{\sigma}}_{\bm{R}+\bm{a}_3}),
\label{eq:Q}
\end{equation}
with $\hat{\vec{\sigma}}_{\bm{r}}=(\hat{\sigma}^x_{\bm{r}},\hat{\sigma}^y_{\bm{r}})$, the mean-field Hamiltonian reads
\begin{equation}
  {\cal H}^{\mathrm{P}}_{\mathrm{MF}}
  =\sum_{\bm{q}}\langle\hat{Q}_{\bm{q}}^\dagger\rangle h^{\mathrm{q}}_{\bm{q}}\langle\hat{Q}_{\bm{q}}\rangle,
  \label{eq:H^q_MF}\\
\end{equation}
where
\begin{widetext}
\begin{equation}
  h^{\mathrm{P}}_{\bm{q}}=2J_{\mathrm{n.n.}}\left(
\begin{array}{cccc}
  0 & (\delta\hat{\tau}_0+q\frac{-\sqrt{3}\hat{\tau}_x-\hat{\tau}_z}{2})f_{q_y+q_z} & (\delta\hat{\tau}_0+q\frac{\sqrt{3}\hat{\tau}_x-\hat{\tau}_z}{2})f_{q_z+q_x} & (\delta\hat{\tau}_0+q\hat{\tau}_z)f_{q_x+q_y}\\
  (\delta\hat{\tau}_0+q\frac{-\sqrt{3}\hat{\tau}_x-\hat{\tau}_z}{2})f_{q_y+q_z} & 0 & (\delta\hat{\tau}_0+q\hat{\tau}_z)f_{q_x-q_y} & (\delta\hat{\tau}_0+\frac{\sqrt{3}\hat{\tau}_x-\hat{\tau}_z}{2})f_{q_z-q_x}\\
  (\delta\hat{\tau}_0+q\frac{\sqrt{3}\hat{\tau}_x-\hat{\tau}_z}{2})f_{q_z+q_x} & (\delta\hat{\tau}_0+q\hat{\tau}_z)f_{q_x-q_y} & 0 & (\delta\hat{\tau}_0+\frac{-\sqrt{3}\hat{\tau}_x-\hat{\tau}_z}{2})f_{q_y-q_z}\\
  (\delta\hat{\tau}_0+q\hat{\tau}_z)f_{q_x+q_y} & (\delta\hat{\tau}_0+\frac{\sqrt{3}\hat{\tau}_x-\hat{\tau}_z}{2})f_{q_z-q_x} & (\delta\hat{\tau}_0+\frac{-\sqrt{3}\hat{\tau}_x-\hat{\tau}_z}{2})f_{q_y-q_z} & 0 
\end{array}\right).\ \ \ \ \ 
\end{equation}
\end{widetext}
We have introduced the Fourier component $Q_{\bm{q}}$ of $Q_{\bm{R}}$ in an analogy to Eq.~\eqref{eq:d_q}.

For $\delta>(|q|+1)/2$, the lowest eigenvalue of $h^{\mathrm{P}}_{\bm{q}}$ is given by 
\begin{equation}
  \varepsilon^{\mathrm{PAF}}_{\bm{q}}=-2J_{\mathrm{n.n.}}(\delta+2|q|)
\label{eq:epsilon^PAF_q}
\end{equation}
for the planar antiferro-pseudospin (PAF) states at the rods $\bm{q}=\frac{2\pi}{a}h(1,\pm1,\pm1)$ with $h$ being an arbitrary real number. It has the 120$^\circ$ planar pseudospin structure within a tetrahedron, which is expressed as
\begin{equation}
  {}^t\langle\hat{Q}_{\bm{q}}\rangle=\left\{\begin{array}{l}
  \left(\vec{0},\vec{n}'_{\bm{a}_0,\bm{a}_1},\vec{n}'_{\bm{a}_0,\bm{a}_2},\vec{n}'_{\bm{a}_0,\bm{a}_3}\right)
 \ \mbox{for $\bm{q}=q_x(1,1,1)$}
\\
  \left(\vec{n}'_{\bm{a}_0,\bm{a}_1},\vec{0},\vec{n}'_{\bm{a}_1,\bm{a}_2},\vec{n}'_{\bm{a}_1,\bm{a}_3}\right)
  \ \mbox{for $\bm{q}=q_x(-1,1,1)$}
\\
  \left(\vec{n}'_{\bm{a}_0,\bm{a}_2},\vec{n}'_{\bm{a}_1,\bm{a}_2},\vec{0},\vec{n}'_{\bm{a}_2,\bm{a}_3}\right)
  \ \mbox{for $\bm{q}=q_x(1,-1,1)$}
  \\
  \left(\vec{n}'_{\bm{a}_0,\bm{a}_3},\vec{n}'_{\bm{a}_1,\bm{a}_3},\vec{n}'_{\bm{a}_2,\bm{a}_3},\vec{0}\right)
  \ \mbox{for $\bm{q}=q_x(1,1,-1)$}
  \end{array}\right.
\label{eq:structure^PAF_q}
\end{equation}
for $q>0$. When $q<0$, $\vec{n}'$ in Eq.~\eqref{eq:structure^PAF_q} are replaced by $\vec{n}$.

In the other case of $\delta<(|q|+1)/2$, the lowest eigenvalue of $h^{\mathrm{P}}_{\bm{q}}$ is given by
\begin{equation}
  \varepsilon^{\mathrm{PF}}_{\bm{q}}=6J_{\mathrm{n.n.}}\delta
  \label{eq:epsilon^PF_q}
\end{equation}
for the planar ferro-pseudospin (PF) state at $\bm{q}=\bm{0}$ and symmetry-related $\bm{q}$ vectors connected by $\frac{4\pi}{a}(n_1,n_2,n_3)$ and/or $\frac{2\pi}{a}(1,1,1)$ with integers $n_1$, $n_2$, and $n_3$. This state has eigenvectors showing a collinear ferroic alignment of the planar components of the pseudospins. 

Because of the saturation of each ordered moment, except at one site per tetrahedron in the case of the PAF phase, these states can be stabilized as the ground state if the energy is lower than in the dipolar state.

\subsection{Mean-field phase diagram}
\label{sub:MFT:phasediagram}

\begin{figure}
\begin{center}
\includegraphics[width=\columnwidth]{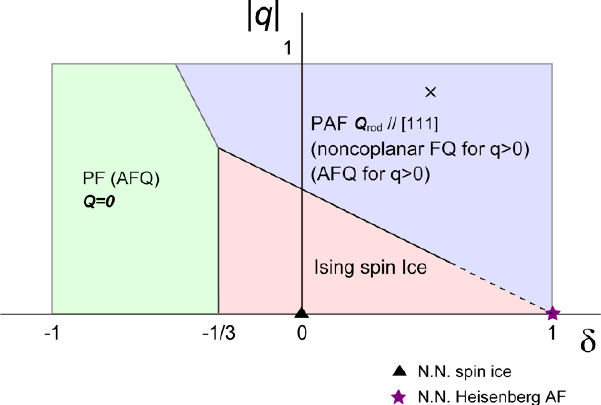}
\end{center}
\caption{(Color online) Classical mean-field phase diagram of the model given by Eq.~\eqref{eq:H_eff}. The $\bm{Q}=\bm{0}$ planar ferro-pseudospin (PF) phase physically represents the antiferroquadrupole (AFQ) phase. The planar antiferro-pseudospin (PAF) phase is characterized by Bragg rods $\bm{Q}\parallel[111]$ and physically represents a 2D AFQ phase for $q>0$ or a 2D noncoplanar ferroquadrupole (FQ) phase for $q<0$, where quadrupole moments are aligned only within the plane perpendicular to Brag rod vectors $\bm{Q}\parallel[111]$ in the mean-field level. In the region around the N.N. Heisenberg antiferromagnet, the present mean-field theory becomes less accurate. The point $X=(\delta,q)=(0.51,0.89)$ for Pr$^{3+}$ is also shown.}
\label{fig:MFT:diagram}
\end{figure}
Comparing the energies of the dipolar state and the two quadrupolar states given above, we obtain the classical mean-field phase diagram shown in Fig.~\ref{fig:MFT:diagram}; (i) the macroscopically degenerate dipolar states associated with the nearest-neighbor spin ice for $-1/3<\delta<1-2|q|$, (ii) the planar antiferro-pseudospin (PAF) states showing the 120$^\circ$ structure of $\langle\hat{\vec{\sigma}}_{\bm{r}}\rangle$ in each plane perpendicular to the rods $\bm{q}=\frac{2\pi}{a}h(1,\pm1,\pm1)$ for $\delta>1-2|q|$, and (iii) a planar ferro-pseudospin (PF) state characterized by $\langle\hat{\bm{\sigma}}_{\bm{r}}\rangle=(\cos\Theta,\sin\Theta,0)$ for $\delta<-1/3$ and $\delta<1-|q|/2$ with an arbitrary angle $\Theta$. Note that the degeneracy along the rods $\bm{q}=q(1,\pm1,\pm1)/\sqrt{3}$ in the PAF phase could be lifted by an order-by-disorder mechanism which favors the $\bm{q}=0$ order because of the higher degeneracy, or by a longer-range interaction which is not taken into account in the present paper. The nearest-neighbor spin ice $(\delta,q)=(0,0)$ and Heisenberg antiferromagnet $(\delta,q)=(1,0)$, both of which are marked in Fig.~\ref{fig:MFT:diagram}, show no LRO~\cite{moessner:98,isakov:04} down to $T=0$ but dipolar spin correlations~\cite{isakov:04}. We stress again that the present mean-field treatment becomes inaccurate around the nearest-neighbor Heisenberg antiferromagnet. Note that the recently studied $120^\circ$ antiferromagnetic planar model~\cite{chern:10} corresponds to the limit of $\delta=|q|\to\infty$.

For Pr$^{3+}$ ions, the planar components $\hat{\vec{\sigma}}_{\bm{r}}$ represent atomic quadrupole moments, as explained in Sec.~\ref{sec:model:moment}. Note that they are defined in local coordinate frames through $\hat{\vec{\sigma}}_{\bm{r}}=((\hat{\bm{\sigma}}_{\bm{r}}\cdot\bm{x}_{\bm{r}}), (\hat{\bm{\sigma}}_{\bm{r}}\cdot\bm{y}_{\bm{r}}))$.  Then, the phases (ii) PAF and (iii) PF are characterized by the following quadrupole order. Since the local frames satisfy Eq.~\eqref{eq:sum_xyz}, the collinear PF state has a noncoplanar antiferroquadrupole LRO without any translation-symmetry breaking. On the other hand, the 2D PAF state has a noncollinear alignment of atomic quadrupole moments in each $[111]$ layer. It exhibits a noncoplanar ferroquadrupole (FQ) order having a finite uniform quadrupole moment pointing to the direction of $\bm{q}$ for $q<0$ or a coplanar 120$^\circ$ antiferroquadrupole order for $q>0$. These can be directly shown by using Eqs.~\eqref{eq:xyz} and \eqref{eq:structure^PAF_q} with or without the replacement of $\vec{n}'$ by $\vec{n}$ for $q<0$ or $q>0$, respectively.

\section{Single-tetrahedron analysis}
\label{sec:single}

It is instructive to investigate the quantum interplay of $\hat{\sigma}^z_{\bm{r}}$ and $\hat{\sigma}^{x,y}_{\bm{r}}$ in the model given by Eq.~(\ref{eq:H_eff}) within a single tetrahedron. A similar analysis on a model for Tb$_2$Ti$_2$O$_7$~\cite{gardner:99} has been employed~\cite{molavian:07}. 

In the classical case of $\beta=\gamma=0$ and thus $\delta=q=0$, there appear three energy levels in a single tetrahedron:
\begin{itemize}
  \item 
    Sixfold degenerate ``2-in, 2-out'' configurations $|\pm X\rangle$, $|\pm Y\rangle$, and $|\pm Z\rangle$ have the energy $-2J_{\mathrm{n.n.}}$ and are characterized by the direction of the net Ising moment on the tetrahedron $T$,
\begin{equation}
  \hat{\bm{{\cal M}}}_T=\sum_{\bm{r}\in T}\hat{\bm{m}}_{\bm{r}}=M_0\sum_{\bm{r}\in T}\hat{\sigma}_{\bm{r}}^z\bm{z}_{\bm{r}},
  \label{eq:M_T}
\end{equation}
 which points to $\pm \bm{X}$, $\pm \bm{Y}$, and $\pm \bm{Z}$ directions in the global frame, respectively. Here, $M_0$ is the local moment amplitude introduced in Eq.~\eqref{eq:M_0}.
\item 
  Eightfold degenerate ``3-in, 1-out'' and ``1-in, 3-out'' configurations have the energy $0$.

\item ``4-in'' and ``4-out'' configurations $|4+\rangle$ and $|4-\rangle$ have the energy $6J_{\mathrm{n.n.}}$
\end{itemize}

With nonzero $\beta$ and $\gamma$ and thus nonzero $\delta$ and $q$, the Hamiltonian Eq.~\eqref{eq:H_eff} can be diagonalized on a single tetrahedron to yield the following set of eigenvalues and eigenstates for a singlet, three doublets, and three triplets;
\begin{enumerate}
\item[(i)] an $A_{1g}$ singlet which is a superposition of the six ``2-in, 2-out'' configurations~\cite{onoda:09,molavian:07};
\begin{subequations}
\begin{eqnarray}
  E_{A_{1g}}&=& -2J_{\mathrm{n.n.}}(1-4\delta),
  \label{eq:E_A1g}\\
  |\Psi_{A_{1g}}\rangle&=&\frac{1}{\sqrt{6}}\sum_{\tau=\pm}\left(|\tau X\rangle+|\tau Y\rangle+|\tau Z\rangle\right).
  \label{eq:A1g}
\end{eqnarray}
\end{subequations}
\item[(ii)] two $E_g$ doublets which are superpositions of both ``2-in, 2-out'' and ``4-in''/``4-out'' configurations~\cite{onoda:09};
\begin{subequations}
\begin{eqnarray}
  E_{E_g}&=& -2 J_{\mathrm{n.n.}}\left(\sqrt{(2+\delta)^2+6 q^2}-1+\delta\right),
  \label{eq:E_Eg}\\
  |\Psi^\chi_{E_g}\rangle&=&\frac{c}{\sqrt{6}}\sum_{\tau=\pm}\left(e^{i\frac{2\pi}{3}\chi}|\tau X\rangle+e^{-i\frac{2\pi}{3}\chi}|\tau Y\rangle+|\tau Z\rangle\right)
  \nonumber\\
  &&{}+c'|4\chi\rangle,
  \label{eq:Eg}
\end{eqnarray}
\end{subequations}
and
\begin{subequations}
\begin{eqnarray}
  &&E_{E_g}'= 2 J_{\mathrm{n.n.}}\left(\sqrt{(2+\delta)^2+6 q^2}+1-\delta\right),
  \label{eq:E_Eg'}\\
&&\frac{c'}{\sqrt{6}}\sum_{\tau=\pm}\left(e^{i\frac{2\pi}{3}\chi}|\tau X\rangle+e^{-i\frac{2\pi}{3}\chi}|\tau Y\rangle+|\tau Z\rangle\right)
-c|4\chi\rangle,
\nonumber\\
  \label{eq:Eg'}
\end{eqnarray}
\end{subequations}
with a sign $\chi=\pm$ and dimensionless functions
  $c'=[\sqrt{6}q/2][(2+d)^2+6q^2+(2+d)\sqrt{(2+d)^2+6q^2}]^{-1}$
and $c=\sqrt{1-c'^2}$,
\item[(iii)] a $T_{1u}$ triplet described with the antisymmetric superposition of ``2-in, 2-out'' configurations
\begin{subequations}
\begin{eqnarray}
  &&E_{T_{1u}}= -2J_{\mathrm{n.n.}},
  \label{eq:E_T1u}\\
&&\frac{1}{\sqrt{2}}\sum_{\tau=\pm}\tau\left(|\tau X\rangle, |\tau Y\rangle, |\tau Z\rangle \right),
  \label{eq:T1u}
\end{eqnarray}
\end{subequations}
and
\item[(iv)] two triplets and a doublet purely comprised of ``3-in, 1-out'' and ``1-in, 3-out'' configurations, whose energy levels are given by $-2J_{\mathrm{n.n.}}(\delta\pm2q)$ and $6J_{\mathrm{n.n.}}\delta$, respectively.

\end{enumerate}

\begin{figure}[tbh]
\begin{center}
\includegraphics[width=\columnwidth]{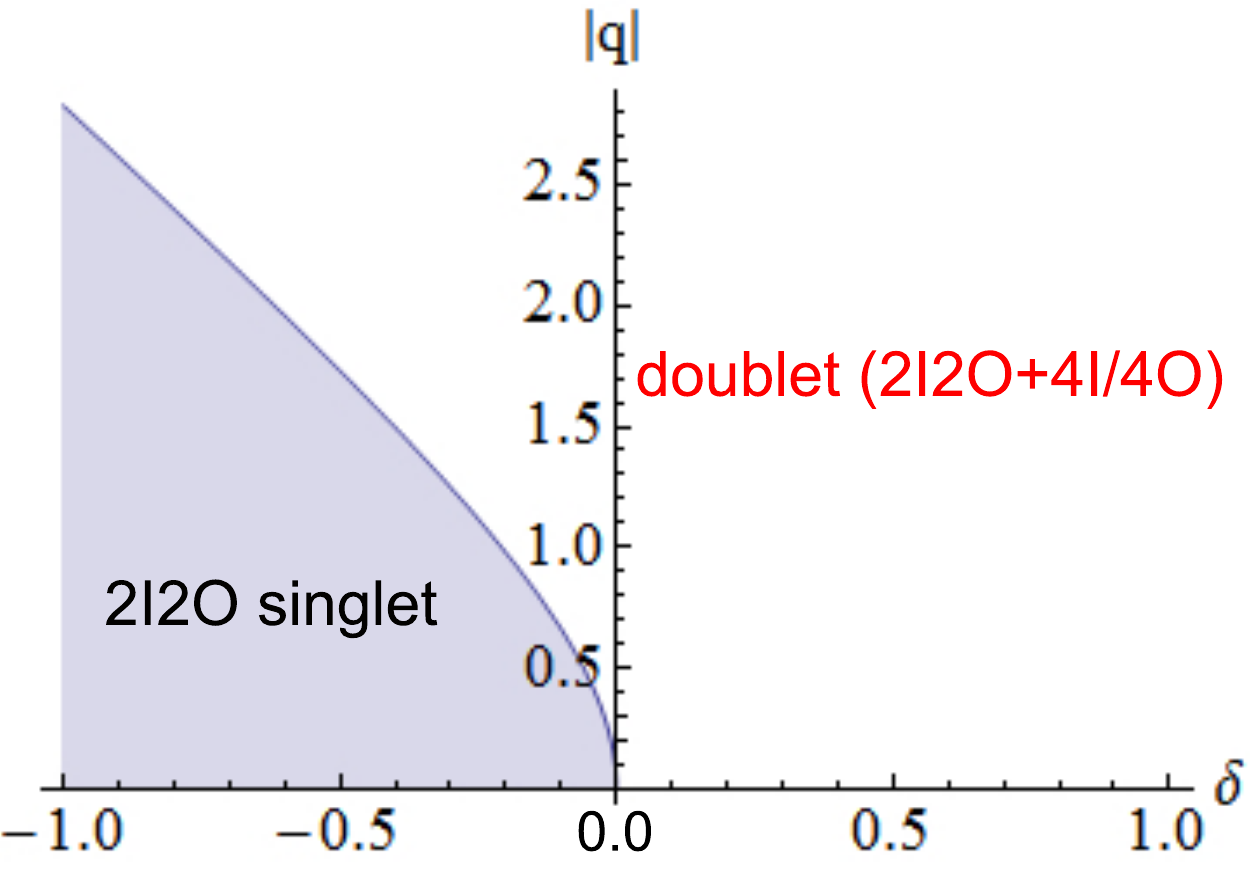}
\end{center}
\caption{(Color online) The symmetry of the ground states of the effective Hamiltonian ${\cal H}_{\mathrm{eff}}$ in the space of $\delta$ and $q$ in a single-tetrahedron analysis. }
\label{fig:single}
\end{figure}

In our case of $J_{\mathrm{n.n.}}>0$, the ground state is given by either a singlet $|\Psi_{A_{1g}}\rangle$ (Eq.~\eqref{eq:A1g}) or a doublet $|\Psi^\chi_{E_g}\rangle$ (Eq.~\eqref{eq:Eg}), depending on whether $\delta$ is less or greater than 
\begin{equation}
  \delta_B(q)=-(\sqrt{1+q^2}-1)/2,
  \label{eq:delta_B}
\end{equation}
respectively, as shown in Fig.~\ref{fig:single}.

\begin{figure}[tbh]
\begin{center}
\includegraphics[width=\columnwidth]{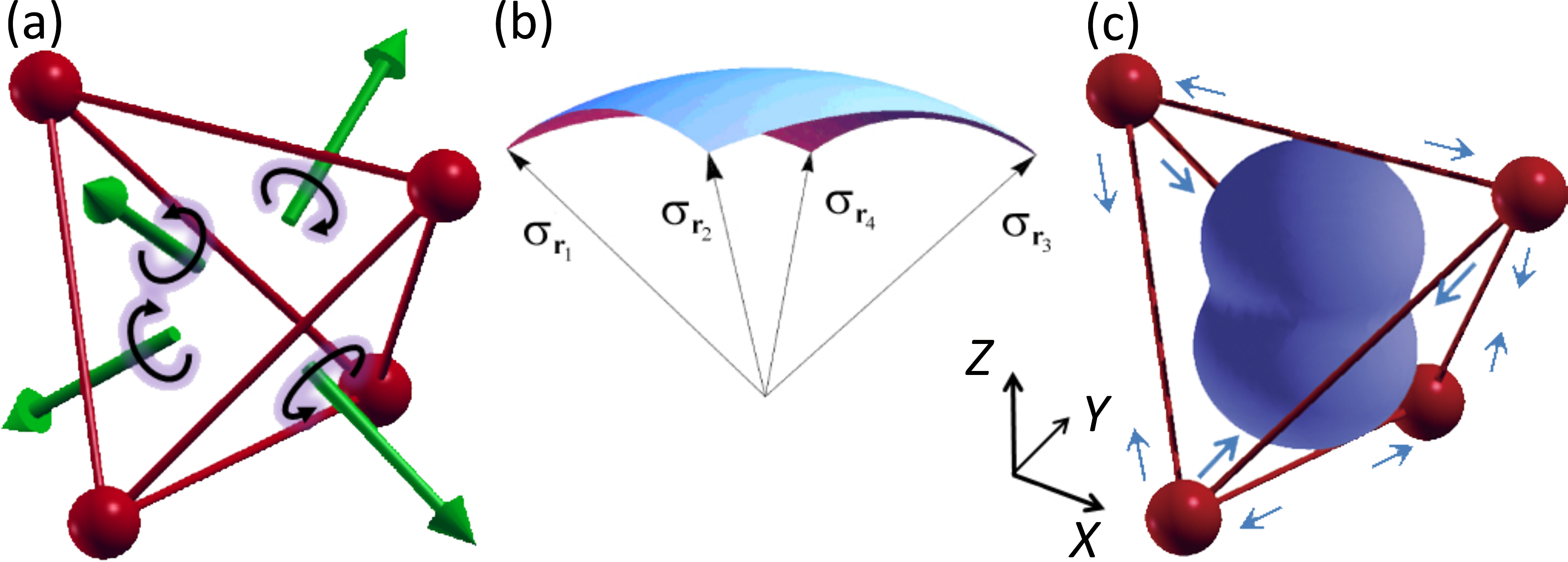}
\end{center}
\caption{(Color online) (a) Outward normal vectors (green arrows) of the surfaces of the tetrahedron, used to define the chirality $\kappa_T$. (b) Solid angle subtended by four pseudospins $\bm{\sigma}_{\bm{r}_i}$. (c) Distribution of the tetrahedral magnetic moment $\bm{{\cal M}}_T$ in a cooperative ferroquadrupolar state with $\langle Q_T^{zz}\rangle>0$. The arrows represent the lattice deformation linearly coupled to $Q_T^{zz}$.}
\label{fig:chiral-quadrupole}
\end{figure}

In the above chiral representation of the doubly degenerate $E_g$ states, $|\Psi_{E_g}^\chi\rangle$, the index $\chi=\pm$ represents the sign of the net pseudospin chirality of the tetrahedron,
\begin{equation}
  \hat{\kappa}_T=\frac{1}{2}\sum_{\bm{r}_1,\bm{r}_2,\bm{r}_3}^T\
  \hat{\bm{\sigma}}_{\bm{r}_1}\cdot\hat{\bm{\sigma}}_{\bm{r}_2}\times\hat{\bm{\sigma}}_{\bm{r}_3},
  \label{eq:kappa}
\end{equation}
through the relation 
\begin{equation}
  \langle\Psi^\chi_{E_g}|\hat{\kappa}_T|\Psi^{\chi'}_{E_g}\rangle=\frac{\sqrt{3}}{2}c^2\chi\delta_{\chi,\chi'}.
  \label{eq:kappa_ave}
\end{equation}
Here, the summation over the sites $\bm{r}_1,\bm{r}_2,\bm{r}_3$ on the tetrahedron $T$ is taken so as they appear counterclockwise about the outward normal to the plane spanned by the three sites for each triangle [Fig.~\ref{fig:chiral-quadrupole} (a)]. $\langle\hat{\kappa}_T\rangle$ gives the solid angle subtended by the four pseudospins [Fig.~\ref{fig:chiral-quadrupole} (b)].

In this level of approximation, the ``2-in, 2-out'' and ``4-in''/''4-out'' configurations are totally decoupled from the ``3-in, 1-out''/''1-in, 3-out'', because any single pseudospin cannot be flipped by the Hamiltonian Eq.~\eqref{eq:H_eff} within a single tetrahedron. This decoupling is an artifact and a drawback of the single-tetrahedron analysis, which is resolved by larger system-size calculations in the next section.

\section{Numerics on the 16-site cube}
\label{sec:ED}

\begin{figure}
\begin{center}
\includegraphics[width=\columnwidth]{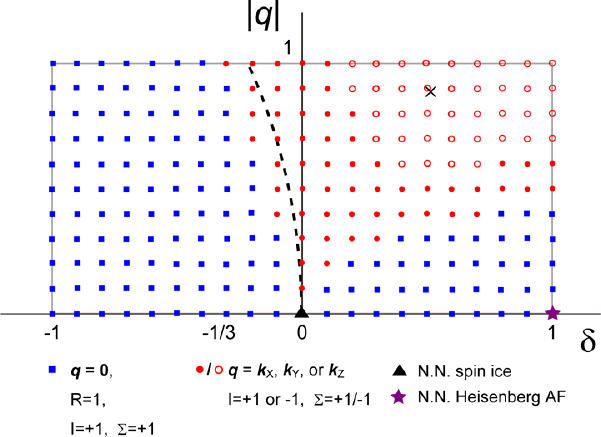}
\end{center}
\caption{(Color online) The symmetry properties of the ground states obtained with the exact diagonalization of the cube in the periodic boundary condition. The dashed curve is the boundary between the ``2-in, 2-out'' singlet and the ``2-in, 2-out''+''4-in/4-out'' doublet for the ground state of the model on a single tetrahedron, as shown in Fig.~\ref{fig:single}. The point $X=(\delta,q)=(0.51,0.89)$ for Pr$^{3+}$ is also shown.}
\label{fig:diagram}
\end{figure}
Next, we perform exact-diagonalization calculations of the model given by Eq.~(\ref{eq:H_eff}) on the 16-site cube [Fig.~\ref{fig:pyrochlore}] in the periodic boundary condition. Because of the lack of the total pseudospin conservation, this system size already gives a large Hilbert space, though we can exploit the following symmetry operations; 
\begin{enumerate}
\item[(i)]
the even-odd parity of the total pseudospin, 
$\hat{\Sigma}=\prod_{\bm{r}}\hat{\sigma}^z_{\bm{r}}$,
\item[(ii)]
the translations $\hat{T}(\bm{R})$ by fcc lattice vectors $\bm{R}=\sum_{i=1,2,3}n_i\bm{R}_i$ with integers $n_i$, 
\item[(iii)]
the spatial inversion $\hat{I}$ about a site, 
\item[(iv)]
the threefold rotation $\hat{R}$ about a $(111)$ axis.
\end{enumerate}
Among these symmetry operations, $\hat{\Sigma}$, $\hat{I}$ and $\hat{R}$ commute with each other. In general, $\hat{T}(\bm{R})$ commutes with $\hat{I}_\sigma$ and $\hat{\Sigma}$ but not with $\hat{I}$ and $\hat{R}$. In our case of the 16-site cube with the periodic boundary condition, however, there exist only two nonequivalent translations $\hat{T}(\bm{R}_i)$ with the fcc primitive lattice vectors $\bm{R}_i$ ($i=1,2$), since $\hat{T}(\bm{R}_1)\hat{T}(\bm{R}_2)=\hat{T}(\bm{R}_2)\hat{T}(\bm{R}_1)=\hat{T}(\bm{R}_3)$ in this case, and these two translations eventually commute with $\hat{I}$. Therefore, we can adopt the following set of commuting operators, $\hat{H}_{\mathrm{eff}}$, $\hat{\Sigma}$, $\hat{T}(\bm{R}_1)$, $\hat{T}(\bm{R}_2)$, and $\hat{I}$. $\hat{R}$ can also be used only in the translationally invariant manifold where both $\hat{T}(\bm{R}_1)$ and $\hat{T}(\bm{R}_1)$ have the eigenvalue $1$. 

In Fig.~\ref{fig:diagram}, we show the symmetry properties of the ground state in the parameter space of $\delta$ and $|q|$. Special points corresponding to the nearest-neighbor spin ice and the nearest-neighbor Heisenberg antiferromagnet are denoted as a triangle and a star, respectively, where dipolar spin correlations appear without any LRO~\cite{moessner:98,isakov:04}. At finite $\delta$ and/or $q$, there appear four regions in the parameter range $\delta, |q|\le1$. The boundary $\delta=\delta_B(q)$ [Eq.~\eqref{eq:delta_B}] between the $A_{1g}$ singlet and the $E_g$ doublet ground states in the single-tetrahedron level, which is shown as the dashed curve, almost gives one of the boundaries. On the left-hand side of the curve, i.e., $\delta\lesssim\delta_B(q)$, the ferroic pseudospin exchange ($\delta<0$) stabilizes a rotationally and translationally invariant singlet ground state with the even parity $I=\Sigma=+1$ for both the spatial inversion and pseudospin parity (filled blue squares). The mean-field result of the collinear ferroic LRO of the planar components of pseudospins should be realized when $\delta$ is negatively large. Therefore, it is plausible to assign most of this region to the PF (AFQ) phase. The ground state having the same symmetry appears in the case of antiferroic pseudospin exchange coupling $\delta>0$ when $|q|$ is much less than $\delta$. Noting that the U(1) spin liquid is stable against a weak antiferroic pseudospin exchange interaction~\cite{hermele:04} and the mean-field result also gives a macroscopically degenerate spin-ice state, this could be assigned to a U(1) spin liquid~\cite{hermele:04} or a quantum spin ice without magnetic dipole LRO. It remains open whether it is also stable against a weak ferroic pseudospin exchange coupling, namely, whether the U(1) spin liquid might appear even in the case of $\delta\lesssim\delta_B(q)$.

On the other hand, increasing $|q|$ out of the above two regions changes the ground state from the singlet to sextets. The sixfold degeneracy of the ground states are described by a product of (i) the double degeneracy characterized by the eigenvalues $+1$/$-1$ for the spatial inversion $\hat{I}$ and (ii) the threefold degeneracy characterized by the three sets of eigenvalues, $(1,-1,-1)$, $(-1,1,-1)$, and $(-1,-1,1)$, for the translation $(\hat{T}(\bm{R}_1),\hat{T}(\bm{R}_2),\hat{T}(\bm{R}_3))$, or equivalently, the wavevectors $\bm{k}_X=\frac{2\pi}{a}(1,0,0)$, $\bm{k}_Y=\frac{2\pi}{a}(0,1,0)$, and $\bm{k}_Z=\frac{2\pi}{a}(0,0,1)$. In fact, the singlet-sextet transition occurs in two steps. The singlet ground state is first replaced by the submanifold of the above sixfold degenerate states that has an even pseudospin-parity $\Sigma=+1$ (filled red circles). With further increasing $|q|$, it is replaced by the other submanifold that has an odd pseudospin-parity $\Sigma=-1$ (open red circles). Though these states might have an antiferroic LRO of planar components of pseudospins as obtained in the mean-field approximation in Sec.~\ref{sec:MFT}, the determination of a possible LRO in these regions is nontrivial within calculations on a small system size. The particular case of $\delta=0.51$ and $q=0.89$ which we have found for Pr$^{3+}$ is also located in the region of the six-fold degenerate ground state, as shown in Fig.~\ref{fig:diagram}, with the energy $\sim -8.825J_{\mathrm{n.n.}}$ per tetrahedron. In the rest of this paper, we will investigate magnetic dipole, quadrupole, and chiral correlations in this particular case.

\subsection{Magnetic dipole correlation}
\label{sec:ED:dipole}

\begin{figure}[t]
\begin{minipage}{0.45\linewidth}
\includegraphics[width=\columnwidth]{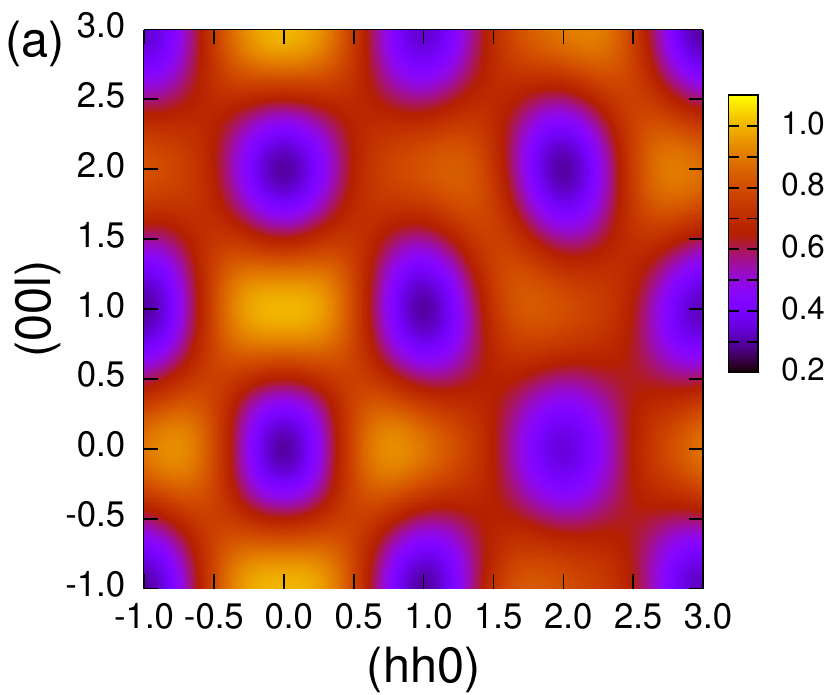}
\end{minipage}
\begin{minipage}{0.45\linewidth}
\begin{center}
\includegraphics[width=\columnwidth]{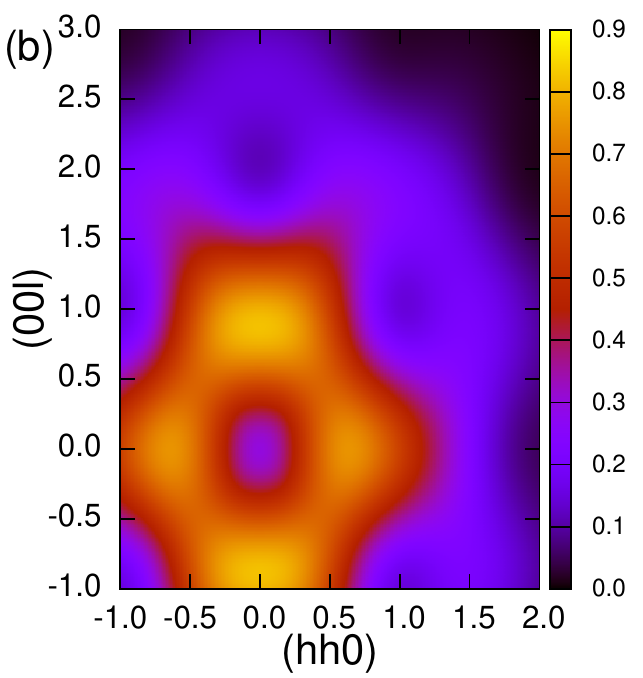}
\end{center}
\end{minipage}
\caption{(Color online) Calculated neutron scattering profile (a) $S(\bm{q})/M_0^2$ and (b) $S(\bm{q})F_{\mathrm{Pr}^{3+}}(\bm{q})^2/M_0^2$ for $\bm{q}=\frac{2\pi}{a}(hhl)$, with the form factor $F_{\mathrm{Pr}^{3+}}(\bm{q})$.
}
\label{fig:Neutron}
\end{figure}

First, we calculate the magnetic dipole correlation,
\begin{equation}
  S(\bm{q})=\frac{M_0^2}{N}\sum_{\bm{r},\bm{r}}\sum_{i,j}(\delta_{ij}-\frac{q_iq_j}{|\bm{q}|^2})z_{\bm{r}}^i z_{\bm{r}'}^j\langle\hat{\sigma}^z_{\bm{r}}\hat{\sigma}^z_{\bm{r}'}\rangle_{\mathrm{ave}}e^{i\bm{q}\cdot(\bm{r}-\bm{r}')},
\end{equation}
averaged over the degenerate ground states. For non-Kramers ions such as Pr$^{3+}$, this quantity is relevant to the neutron-scattering intensity integrated over the low-energy region below the crystal-field excitations from the atomic ground-state doublet Eq.~(\ref{eq:local}), while for Kramers ions, the transverse components $\hat{\sigma}_{\bm{r}}^{x,y}$ must also be taken into account.
In Fig.~\ref{fig:Neutron}, we show the profiles of (a) $S(\bm{q})/M_0^2$ and (b) $S(\bm{q})/M_0^2\cdot F_{\mathrm{Pr}^{3+}}(|\bm{q}|)^2$ for $\bm{q}=\frac{2\pi}{a}(hhl)$ with $F_{\mathrm{Pr}^{3+}}(q)$ being the form factor for Pr$^{3+}$. It exhibits maxima at $(001)$ and $(003)$ as well as at $(\frac{3}{4}\frac{3}{4}0)$, and a minimum at $(000)$, as observed in the dipolar spin ice~\cite{bramwell:01}. Note however that the calculated profiles are constructed from the on-site, nearest-neighbor, and second-neighbor correlations, which gives a good approximation when the magnetic dipole correlations remain short-range. Obviously, when the spin correlation length is longer, we need calculations on a larger system size, in particular, around the wavevectors such as (111) and (002), where the pinch-point singularity~\cite{isakov:04,henley:05} appears as in the case of classical spin ice~\cite{fennell:07}. Nevertheless, the failure of the strict ``2-in, 2-out'' ice rule can broaden the singularity.

\begin{figure}[t]
\begin{center}
\includegraphics[width=\columnwidth]{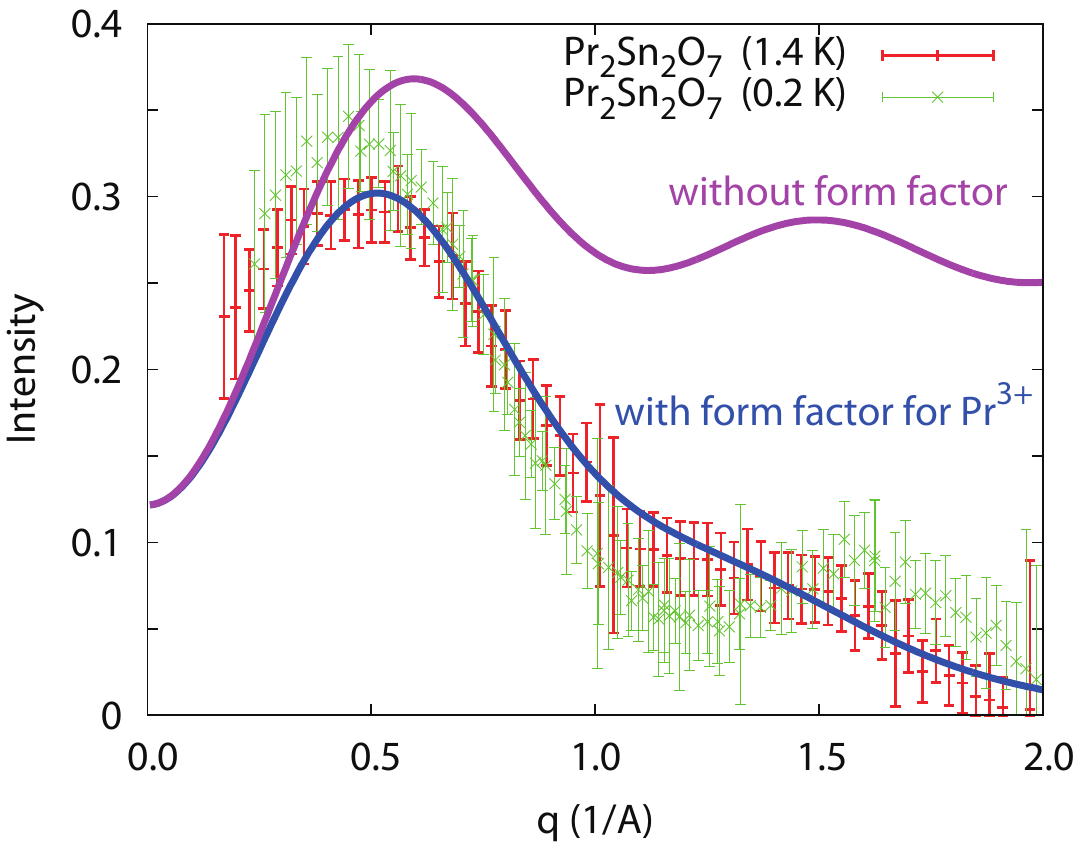}
\end{center}
\caption{(Color online) Poweder neutron-scattering intensity. Theoretical curves with/without the form factor (blue/magneta curve) and the experimental results on Pr$_2$Sn$_2$O$_7$ from Ref.~\onlinecite{zhou:08}.}
\label{fig:PowNeut}
\end{figure}

At the moment, there are not so many experimental results on the magnetic dipole correlations in Pr$_2TM_2$O$_7$. The only currently available one is a powder neutron-scattering experiment on Pr$_2$Sn$_2$O$_7$~\cite{zhou:08}.
It reveals the absence of magnetic Bragg peaks and the enhanced low-energy short-ranged intensity around $|\bm{q}|\sim\frac{2\pi}{a}\sim0.5$~\AA$^{-1}$ and $\sim\frac{6\pi}{a}\sim1.5$~\AA$^{-1}$. 
These features can be explained in terms of the peak positions in our calculated profile which are similar to the dipolar spin ice, as shown in Fig.~\ref{fig:PowNeut}. This experiment also shows the quasielastic peak width $\sim0.1$~meV saturated at 0.2~K~\cite{zhou:08}. Such large dynamical spin relaxation rate $\sim J_{\mathrm{n.n.}}$ can be attributed to the appreciable quantum nature of the Hamiltonian, i.e., large $\delta$ and $q$.

\subsection{Magnetization curve}
\label{sec:ED:MH}

\begin{figure}[tbh]
\begin{center}
\includegraphics[width=\columnwidth]{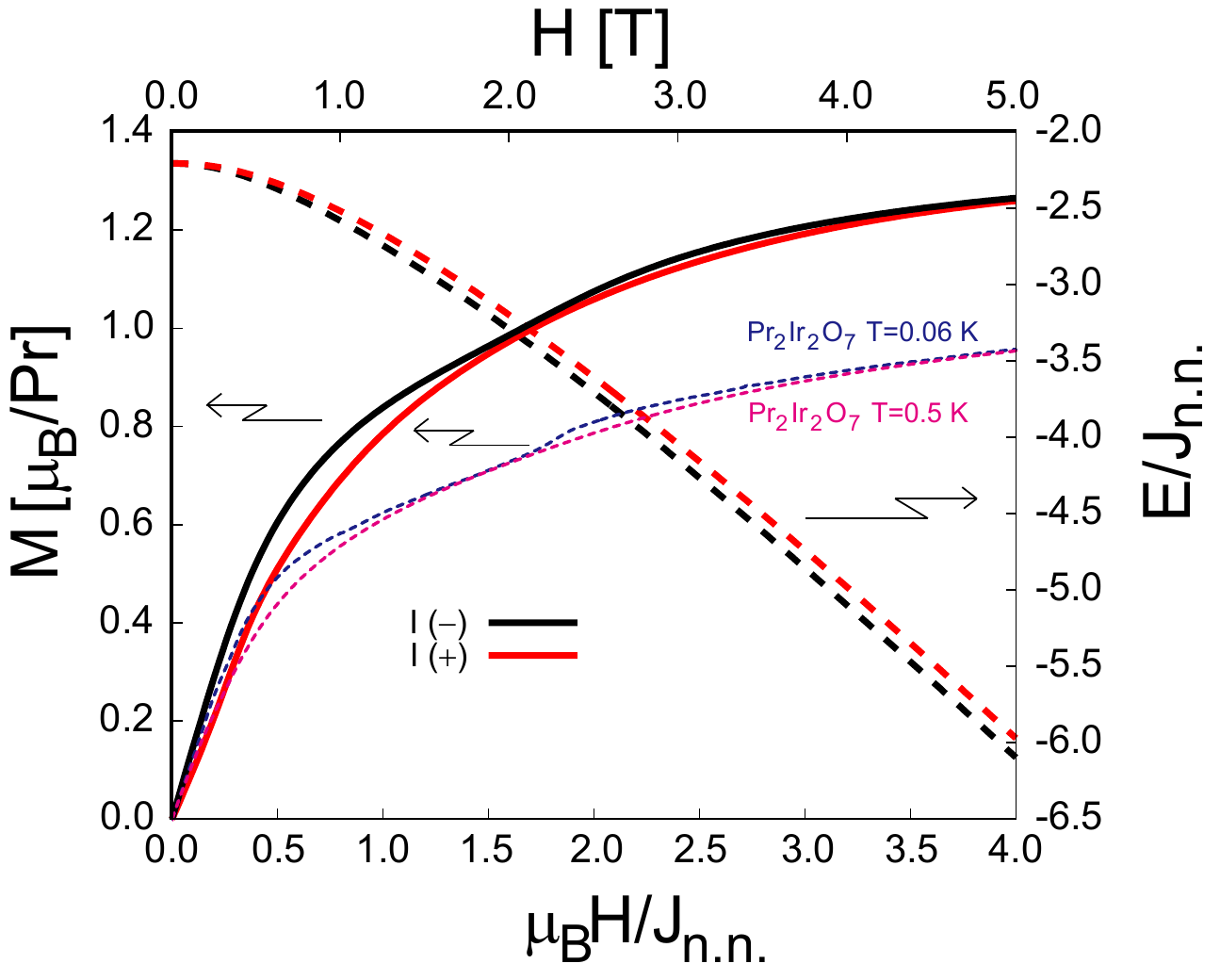}
\end{center}
\caption{(Color online) The magnetization (left) and energy (right) per site calculated for the $I$-odd(-) ground state and the $I$-even(+) state  by adding the Zeeman term $-\bm{H}\cdot\bm{M}$ to ${\cal H}_{\mathrm{eff}}$ at the applied field $\bm{H}\parallel[111]$. The symmetry of the ground-state manifold does not change until a level cross occurs at $\mu_B H/J_{\mathrm{n.n.}}\sim4.6$ to the almost fully polarized state. Experimental data on Pr$_2$Ir$_2$O$_7$ at $T=0.5$ K and 0.06 K from Ref.~\onlinecite{machida:09} are also shown for comparison, though they include additional contributions from Ir conduction electrons.}
\label{fig:MH}
\end{figure}

Next, we show the magnetization curve. The applied magnetic field $\bm{H}\parallel[111]$ partially lifts the ground-state degeneracy associated with the inversion symmetry: it splits the energies of the $I$-odd(-) and $I$-even(+) ground-state manifolds. Then, the ground state has the $I$-odd(-) property. In both submanifolds, the magnetic susceptibility is finite, as seen from the slope of the magnetization curve shown in Fig.~\ref{fig:MH}. This is consistent with a finite magnetic susceptibility and thus the negative $T_{CW}$ in Pr$_2$Zr$_2$O$_7$~\cite{matsuhira:09} and Pr$_2$Ir$_2$O$_7$~\cite{machida:09}, and with the absence of an internal magnetic field in Pr$_2$Ir$_2$O$_7$~\cite{maclaughlin:08}. The ground-state $I$-odd(-) magnetization curve shows a small step or dip around $\mu_BH/J_{\mathrm{n.n.}}\sim1.8$, in comparison with that of the $I$-even(+) excited state. This indicates that this structure develops upon cooling. These results qualitatively agree with the experimental results indicating the absence and the emergence of the metamagnetic transition at $H_c\sim2.3$~T for $T=0.5$~K and 0.06~K, respectively, on Pr$_2$Ir$_2$O$_7$~\cite{machida:09} [Fig.~\ref{fig:MH}]. Requiring that $\mu_B H_c/J_{\mathrm{n.n.}}\sim1.8$, we can estimate the effective ferromagnetic Ising coupling as  $J_{\mathrm{eff}}\sim J_{\mathrm{n.n.}}\sim0.84$~K. However, the magnitude of the magnetization is overestimated by about 25~\%, probably because the experimental results on Pr$_2$Ir$_2$O$_7$ include contributions from Ir conduction electrons. Experiments on the single crystals of insulating compounds Pr$_2$Zr$_2$O$_7$ are not available at the moment but could directly test our theoretical model in the quantitative level.

\subsection{Multipole correlations}

The sixfold degenerate ground state can be written as a linear combination,
\begin{equation}
  |GS\rangle=\sum_{i=X,Y,Z}\sum_{I=\pm}c_{i,I}|\Psi_{i,I}\rangle,
  \label{eq:GS}
\end{equation}
where $c_{i,I}$ being complex constants satisfying the normalization condition $\sum_{i,I}|c_{i,I}|^2=1$. Here, we have introduced a ground state $|\Psi_{i,I}\rangle$ associated with $\bm{k}_i$ for both $I=+1$ and $I=-1$, which shows a finite cooperative quadrupole moment defined on each tetrahedron, $\langle\Psi_{i,I}|\hat{{\cal Q}}^{jj}_T|\Psi_{i,I}\rangle=0.0387M_0^2\delta_{ij}$, where
\begin{equation}
  \hat{{\cal Q}}^{ij}_T=3\hat{{\cal M}}^i_T\hat{{\cal M}}^j_T-\hat{\bm{{\cal M}}}^2_T\delta_{ij},
\mbox{ $i,j=X, Y, Z$ (global axes)}.
\end{equation}
Namely, for instance, in the ground state sector associated with $\bm{q}=\bm{k}_Z$, the net magnetic moment $\bm{{\cal M}}_T$ in each tetrahedron $T$ is inclined to point to the $\pm Z$ directions with a higher probability than to the $\pm X$ and $\pm Y$, as shown in Fig.~\ref{fig:chiral-quadrupole} (c). This reflects a $C_3$ symmetry breaking in the choice of the ground-state sector $\bm{q}=\bm{k}_i$. Thus, if we impose the $C_3$ symmetry to the ground state given by Eq.~\eqref{eq:GS}, it is of course possible to cancel the cooperative quadrupole moment, $\langle GS|\hat{{\cal Q}}^{ij}_T|GS\rangle=0$. However, it is natural to expect that the discrete $C_3$ symmetry is eventually broken in the thermodynamic limit. Therefore, we study properties of the ground state having a particular wavevector $\bm{k}_Z$ and thus a direction for the finite cooperative quadrupole moment $\langle GS|\hat{{\cal Q}}^{ZZ}_T|GS\rangle\ne0$. Note that this state shows not only axial alignments of magnetic dipoles but also a broken translational symmetry, and can then be classified into a magnetic analog of a smectic (or crystalline) phase of liquid crystals~\cite{degenne}.

The cooperative quadrupole moment $\langle\hat{{\cal Q}}_T^{ii}\rangle$ linearly couples to a lattice distortion to be verified experimentally: the four ferromagnetic bonds and the two antiferromagnetic bonds should be shortened and expanded, respectively, leading to a crystal symmetry lowering from cubic to tetragonal accompanied by a compression in the direction of the ferroquadrupole moment [Fig.~\ref{fig:chiral-quadrupole} (c)]. This also indicates that a uniaxial pressure along $[100]$ directions can align possible domains of quadrupole moments. Experimental clarification of magnetic dipole and quadrupole correlations in Pr$_2TM_2$O$_7$ by NMR is intriguing.

\begin{figure}[t]
\begin{center}
\includegraphics[width=\columnwidth]{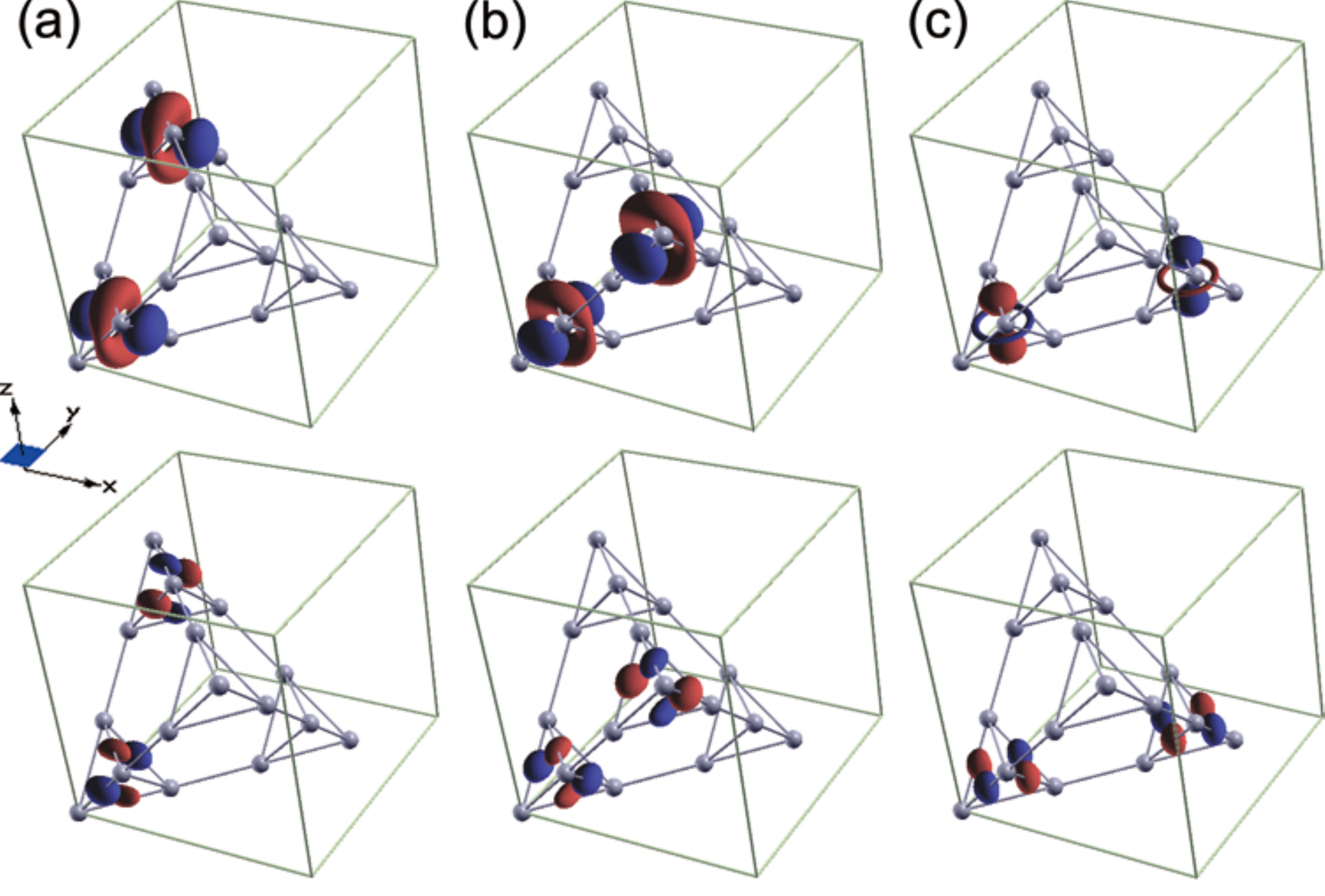}
\end{center}
\caption{
(Color online) Upper/lower panels: dominant/subdominant forms of quadrupole-quadrupole correlations $\langle\hat{{\cal Q}}^{ii}_T\hat{{\cal Q}}^{jj}_{T'}\rangle$ between the tetrahedrons $T$ and $T'$ displaced by $\bm{R}=\bm{R}_1$ (a), $\bm{R}_2$ (b), and $\bm{R}_3$ (c) in the cooperative ferroquadrupolar state with the wavevector $\bm{k}_Z$ and $\langle {\cal Q}_T^{zz}\rangle\ne0$.
 In particular, $\bm{\lambda}_{\bm{R}_1,1}=(-0.803,0.274,0.529)$, $\bm{\lambda}_{\bm{R}_2,1}= (0.274,-0.893,0.529)$, and $\bm{\lambda}_{\bm{R}_3,1}= (1,-1,0)/\sqrt{2}$. Red and blue regions represent the positive and negative values of ${\cal Q}_{\bm{R},1}$.}
\label{fig:q}
\end{figure}
Next, we perform numerical calculations of equal-time spatial correlations of cooperative quadrupole moments, $\langle {\cal Q}^{ii}_T{\cal Q}^{jj}_{T'}\rangle$, between the tetrahedrons $T$ and $T'$ displaced by $\bm{R}$ which take matrix form in $i$ and $j$. To characterize the real-space correlations, we diagonalize this matrix to obtain the two correlation amplitudes,
\begin{equation}
  F^{\cal Q}_{\bm{R},\mu}= \langle\hat{{\cal Q}}_{T,\bm{R},\mu}\hat{{\cal Q}}_{T',\bm{R},\mu}\rangle,
  \label{eq:F^Q_R}
\end{equation}
having orthogonal forms of quadrupoles,
\begin{equation}
  \hat{{\cal Q}}_{T,\bm{R},\mu}=\sum_{i=X,Y,Z}\lambda_{\bm{R},\mu}^i\hat{Q}_T^{ii}, 
  \label{eq:Q_T}
\end{equation}
where $\bm{\lambda}_{\bm{R},\mu}=(\lambda_{\bm{R},\mu}^X,\lambda_{\bm{R},\mu}^Y,\lambda_{\bm{R},\mu}^Z)$ with $\mu=1,2$ is a set of orthonormal vectors satisfying $\sum_{i=X,Y,Z}\lambda_{\bm{R},\mu}^i\lambda_{\bm{R},\nu}^i=\delta_{\mu\nu}$ and $\sum_{i=X,Y,Z}\lambda_{\bm{R},\mu}^i=0$. 
Figures~\ref{fig:q} (a), (b), and (c) show the contour plots of $\lambda_{\bm{R},\mu}^X X^2+\lambda_{\bm{R},\mu}^Y Y^2+\lambda_{\bm{R},\mu}^Z Z^2$, which represent the diagonalized shapes for the quadrupole-quadrupole correlations $\langle \hat{{\cal Q}}^{ii}_T\hat{{\cal Q}}^{jj}_{T'}\rangle$ between the tetrahedrons $T$ and $T'$ displaced by $\bm{R}_1=(0,a/2,a/2)$, $\bm{R}_2=(a/2,0,a/2)$, and $\bm{R}_3=(a/2,a/2,0)$, respectively. Here, red and blue colors represent the positive and negative values. The upper and lower panels correspond to the forms ${\cal Q}_{\bm{R},1}$ and ${\cal Q}_{\bm{R}2}$ showing the larger and smaller correlation amplitudes, respectively. There exist dominant ferroquadrupolar correlations shown in the upper panels of (a) and (b), both of which favors ferroquadrupole moments along the $z$ direction. They prevail over antiferroquadrupole correlations shown in (c), and are responsible for forming the ferroquadrupole order $\langle\hat{{\cal Q}}_T^{zz}\rangle\ne0$. 

To gain insight into a ``chiral spin state'' observed in Pr$_2$Ir$_2$O$_7$, we have also performed numerical calculations of the chirality-chirality correlation $\langle\hat{\kappa}_T\hat{\kappa}_{T'}\rangle$ between the tetrahedrons at $T$ and $T'$. Note that the chirality $\hat{\kappa}_T$ is a pseudospin chirality defined through Eq.~\eqref{eq:kappa}, and is not a simple one defined only with the Ising dipole moments $\hat{\sigma}^z_{\bm{r}}\bm{z}_{\bm{r}}$. It turned out that this pseudospin chirality correlation $\langle\hat{\kappa}_T\hat{\kappa}_{T'}\rangle$ is weakly ferrochiral between the tetrahedrons displaced by $\bm{R}_1$ and $\bm{R}_2$, which corresponds to Figs.~\ref{fig:q} (a) and (b). On the other hand, it is strongly antiferrochiral between those displaced by $\bm{R}_3$, which corresponds to Fig.~\ref{fig:q} (c). Namely, the pseudospin chirality, which is a scalar quantity defined on the tetrahedrons forming a diamond lattice, dominantly shows an antiferrochiral correlation on the nearest-neighbor pairs of the same fcc sublattice of the diamond lattice. This points to a strong geometrical frustration for a chirality ordering. The fate of this pseudospin chirality correlation should be examined by further investigations, which may open an intriguing possibility of a chiral spin liquid~\cite{wen:89}.

\section{Discussions and summary}
\label{sec:summary}

The effective quantum pseudospin-$1/2$ model is quite generically applicable to other pyrochlore magnets associated with rare-earth magnetic moments, though the values of three coupling constants for non-Kramers ions and four for Kramers ions may depend largely on the materials. In this paper, we have concentrated on novel quantum effects in the case of non-Kramers ions, in particular, Pr$^{3+}$ ions, where we expect the most pronounced quantum effects among the rare-earth magnetic ions available for magnetic pyrochlores~\cite{subramanian:83,gardner:10}. The quantum effects may result in two different scenarios, depending on the values of coupling constants: (i) a quantum spin ice where the quantum-mechanical mixing of ``3-in, 1-out'' and ``1-in, 3-out'' could be integrated out to bear quantum effects in magnetic monopoles, and (ii) a ferroquadrupolar state that replaces the spin ice because of a quantum melting. We have obtained ferroquadrupolar state for the case of Pr$^{3+}$, whose magnetic properties explain currently available experimental observations in Pr$_2$Sn$_2$O$_7$ and Pr$_2$Ir$_2$O$_7$. Note that long-distance properties are still beyond the scope of our present calculations on finite-size systems. Further extensive studies from both theoretical and experimental viewpoints are required for the full understanding of nontrivial quantum effects in these systems, in particular, Pr$_2TM_2$O$_7$ and Tb$_2TM_2$O$_7$. Also from a purely theoretical viewpoint, it will be an intriguing and urgent issue to clarify the fate of deconfined magnetic monopoles under the circumstance of large quantum-mechanical interactions we derived.

Effects of coupling of localized $f$-electrons to conduction electrons on the transport properties are left for a future study. A coupling of Pr moments to the atomic and/or delocalized orbital degrees of freedom of conduction electrons allows a flip of the pseudospin-$1/2$ for the magnetic doublet of Pr ion. This could be an origin of the resistivity minimum observed in Pr$_2$Ir$_2$O$_7$~\cite{nakatsuji:06}.

\acknowledgments
The authors thank L. Balents, M. P. Gingras, S. Nakatsuji, C. Broholm, Y. Machida, K. Matsuhira, and D. MacLaughlin for useful discussions. The work was partially supported by Grants-in-Aid for Scientific Research under No. 19052006 from the Ministry of Education, Culture, Sports, Science, and Technology (MEXT) of Japan and No. 21740275 from the Japan Society of Promotion of Science, and by the NAREGI Nanoscience Project from the MEXT. S.O. is also grateful to the hospitality during the stay in Johns Hopkins University where a part of the work was performed.

\appendix
\section{Crystalline Electric Field}
\label{app:CEF}

Here, we give the formal expression for the crystalline electric field (CEF) acting on a Pr site.
In this Appendix, the Pr site is taken as the origin for convenience. The coordinate frame is chosen as that defined in Fig.~\ref{fig:crystal} (b).

We consider the CEF created by nearby ions; two oxygen ions at the O1 sites, six $TM$ ions, and six oxygen ions at the O2 sites.

\subsection{CEF from O1 sites}
\label{app:CEF:O1}

Two oxygen ions at the O1 sites are located at $\pm\frac{\sqrt{3}a}{8}\bm{z}$ and produce the Coulomb potential
\begin{eqnarray}
  \lefteqn{
    U_{O1}(\bm{r}) = q_{O1}\sum_{\tau=\pm}\frac{1}{|\bm{r}-\tau\frac{\sqrt{3}a}{8}\bm{z}|}
    }
  \nonumber\\
  &=&\frac{16q_{O1}}{\sqrt{3}a}\sum_{\ell=0}^\infty\left(\frac{8r}{\sqrt{3}a}\right)^{2\ell}\sqrt{\frac{4\pi}{4\ell+1}}Y_{2\ell}^0(\Omega_{\bm{r}}),
  \label{eq:CEF:O1}
\end{eqnarray}
at a position $\bm{r}$ from the Pr site, where $r=|\bm{r}|$ is assumed to be smaller than $\sqrt{3}{a}/8$, $\Omega_{\bm{r}}$ represents the spherical coordinates of $\bm{r}$, and $q_{O1}\sim-2|e|$ is an effective charge of the oxygen ions at O1 sites.

\subsection{CEF from $TM$ sites}
\label{app:CEF:TM}

Six $TM$ ions are located at $\frac{a}{2}\bm{y}$ and its symmetry related points obtained by successively applying the sixfold rotation $R_6$ about $z$, and produce the Coulomb potential
\begin{eqnarray}
  \lefteqn{
    U_{TM}(\bm{r}) = q_{TM}\sum_{n=0}^5\frac{1}{|\bm{r}-\frac{a}{2}R_6^n\bm{y}|}
    }
  \nonumber\\
  &=&\frac{12q_{TM}}{a}\sum_{\ell=0}^\infty\left(\frac{2r}{a}\right)^{2\ell}\frac{4\pi}{4\ell+1}\sum_{6|m|\le2\ell}Y_{2\ell}^{6m*}(\Omega_{\bm{r}})Y_{2\ell}^{6m}(\Omega_{TM})
  \nonumber\\
  &=&\frac{12q_{TM}}{a}\sum_{\ell=0}^3\left(\frac{2r}{a}\right)^{2\ell}\frac{4\pi}{4\ell+1}Y_{2\ell}^0(\Omega_{\bm{r}})Y_{2\ell}^0(\Omega_{TM})
  \nonumber\\
  &&+\frac{12q_{TM}}{a}\left(\frac{2r}{a}\right)^6\frac{4\pi}{13}\sum_{m=\pm1}Y_6^{6m*}(\Omega_{\bm{r}})Y_6^{6m}(\Omega_{TM})
  \nonumber\\
  &&+\cdots,
  \label{eq:CEF:TM}
\end{eqnarray}
at a position $\bm{r}$ from the Pr site, where $r=|\bm{r}|$ is assumed to be smaller than $a/2$, $\Omega_{TM}=(\frac{\pi}{2},\frac{\pi}{2})$, and $q_{TM}\sim+4|e|$ is an effective charge of the $TM$ ions.

\subsection{CEF from O2 sites}
\label{app:CEF:O2}

Two oxygen ions at the O1 sites are located at $\pm(\sqrt{2}(\frac{1}{8}-\eta)\bm{x}+\eta\bm{z})$ and their symmetry-related points obtained by successively applying the threefold rotation $R_3$ about $\bm{z}$, and produce the Coulomb potential
\begin{eqnarray}
  \lefteqn{
    U_{O2}(\bm{r}) = q_{O2}\sum_{\tau=\pm}\sum_{n=0}^2\frac{1}{|\bm{r}-\tau(\sqrt{2}(\frac{1}{8}-\eta)R_3^n\bm{x}+\eta\bm{z})|}
    }
  \nonumber\\
  &=&\frac{6q_{O2}}{b_{O2}}\sum_{\ell=0}^\infty\left(\frac{r}{b_{O2}}\right)^{2\ell}\frac{4\pi}{4\ell+1}\sum_{3|m|\le2\ell}Y_{2\ell}^{3m*}(\Omega_{\bm{r}})Y_{2\ell}^{3m}(\Omega_{O2}),
  \nonumber\\
  \label{eq:CEF:O2}
\end{eqnarray}
at a position $\bm{r}$ from the Pr site, where $r=|\bm{r}|$ is assumed to be smaller than the Pr-O2 bond length $b_{O2}=\sqrt{3(\frac{1}{32}-\eta/2+3\eta^2)}a$, $\Omega_{O2}=(\theta_2,0)$ with $\theta_2=\arctan(\frac{\sqrt{2}}{\eta}(\frac{1}{8}-\eta))$, and $q_{O2}\sim-2|e|$ is an effective charge of the oxygen ions at O2 sites.

\subsection{Matrix elements between the $f$-electron wavefunctions}
\label{app:CEF:1}

To take into account the orbital dependence of the CEF for $f$ electrons ($l=3$ and $m_l=-3,\cdots,3$), it is sufficient to include the following terms for $U(\bm{r})=U_{O1}(\bm{r})+U_{TM}(\bm{r})+U_{O2}(\bm{r})$;
\begin{eqnarray}
  U(\bm{r})&=&
  \sum_{\ell=1}^3u_{2\ell}^{0*}(r)Y_{2\ell}^0(\Omega_{\bm{r}})
  \nonumber\\
  &&+\sum_{\ell=2}^3\left[u_{2\ell}^{3*}(r)Y_{2\ell}^3(\Omega_{\bm{r}})+u_{2\ell}^{-3*}(r)Y_{2\ell}^{-3}(\Omega_{\bm{r}})\right]
  \nonumber\\
  &&+u_6^{6*}(r)Y_6^6(\Omega_{\bm{r}})+u_6^{-6*}(r)Y_6^{-6}(\Omega_{\bm{r}}),
\end{eqnarray}
where
\begin{subequations}
\begin{eqnarray}
  u_{2\ell}^0(r)&=&\frac{16q_{O1}}{\sqrt{3}a}\left(\frac{8r}{\sqrt{3}a}\right)^{2\ell}\sqrt{\frac{4\pi}{4\ell+1}}
  \nonumber\\
  &&+\frac{12q_{TM}}{a}\left(\frac{2r}{a}\right)^{2\ell}\frac{4\pi}{4\ell+1}Y_{2\ell}^0(\Omega_{TM})
  \nonumber\\
  &&+\frac{6q_{O2}}{b_{O2}}\left(\frac{r}{b_{O2}}\right)^{2\ell}\frac{4\pi}{4\ell+1}Y_{2\ell}^0(\Omega_{O2}),
  \label{eq:u_2l^0}\\
  u_{2\ell}^{\pm3}(r)&=&\frac{6q_{O2}}{b_{O2}}\left(\frac{r}{b_{O2}}\right)^{2\ell}\frac{4\pi}{4\ell+1}Y_{2\ell}^{\pm3}(\Omega_{O2}),
  \label{eq:u_2l^3}\\
  u_6^{\pm6}(r)&=&\frac{6q_{O2}}{b_{O2}}\left(\frac{r}{b_{O2}}\right)^6\frac{4\pi}{13}Y_6^{\pm6}(\Omega_{O2})
  \nonumber\\
  &&+\frac{12q_{TM}}{a}\left(\frac{2r}{a}\right)^6\frac{4\pi}{13}Y_6^{\pm6}(\Omega_{TM}).
\end{eqnarray}
\end{subequations}

Then, their nonvanishing matrix elements of
\begin{equation}
  V_{\mathrm{CEF}}^{m_l,m_l'}=\int\!d\Omega\,Y_3^{m_l*}(\Omega)\overline{U(\bm{r})}Y_3^{m_l'}(\Omega)
  \label{eq:CEF:V_mm'}
\end{equation}
 are obtained as
\begin{subequations}
\begin{eqnarray}
  V_{\mathrm{CEF}}^{m,m}&=&\left\{\begin{array}{r}
  -\frac{1}{6}\sqrt{\frac{5}{\pi}}\bar{u}_2^0+\frac{3}{22\sqrt{\pi}}\bar{u}_4^0-\frac{5}{66\sqrt{13\pi}}\bar{u}_6^0
  \\
  (m=\pm3)
  \\
  -\frac{7}{22\sqrt{\pi}}\bar{u}_4^0+\frac{5}{11\sqrt{13\pi}}\bar{u}_6^0
  \\
  (m=\pm2)
  \\
  \frac{1}{10}\sqrt{\frac{5}{\pi}}\bar{u}_2^0+\frac{1}{22\sqrt{\pi}}\bar{u}_4^0-\frac{25}{22\sqrt{13\pi}}\bar{u}_6^0
  \\
  (m=\pm1)
  \\
  \frac{2}{15}\sqrt{\frac{5}{\pi}}\bar{u}_2^0+\frac{3}{11\sqrt{\pi}}\bar{u}_4^0+\frac{50}{33\sqrt{13\pi}}\bar{u}_6^0
  \\
  (m=0)
  \end{array}\right.,\ \ \ \ \ 
  \label{eq:CEF:V_mm}\\
  V_{\mathrm{CEF}}^{0,3}&=&\left(V_{\mathrm{CEF}}^{3,0}\right)^*=-V_{\mathrm{CEF}}^{-3,0}=-\left(V_{\mathrm{CEF}}^{0,-3}\right)^*
  \nonumber\\
  &=&-\frac{3}{22}\sqrt{\frac{7}{\pi}}\bar{u}_4^3+\frac{5}{11}\sqrt{\frac{7}{39\pi}}\bar{u}_6^3,
  \label{eq:CEF:V_30}
\end{eqnarray}
\begin{eqnarray}
  V_{\mathrm{CEF}}^{-1,2}&=&\left(V_{\mathrm{CEF}}^{2,-1}\right)^*=-V_{\mathrm{CEF}}^{-2,1}=-\left(V_{\mathrm{CEF}}^{1,-2}\right)^*
  \nonumber\\
  &=&-\frac{1}{11}\sqrt{\frac{7}{2\pi}}\bar{u}_4^3-\frac{5}{11}\sqrt{\frac{21}{26\pi}}\bar{u}_6^3,
  \label{eq:CEF:V_2-1}\\
  V_{\mathrm{CEF}}^{-3,3}&=&\left(V_{\mathrm{CEF}}^{3,-3}\right)^*=-5\sqrt{\frac{7}{429\pi}}\bar{u}_6^6,
  \label{eq:VEF:V_6-6}
\end{eqnarray}
  \label{eq:CEF:V}
\end{subequations}
where $\overline{U(\bm{r})}$ and $\bar{u}_{2\ell}^m$ represent the radial averages of $U(\bm{r})$ and $u_{2\ell}^m(r)$.

\section{Representation of the ${}^4H_3$ manifold in terms of single $f$-electrons}
\label{app:f}

\begin{widetext}
\begin{subequations}
\begin{eqnarray}
  |M_J=4\sigma\rangle
  &=&
  \frac{1}{\sqrt{55}}|5,3\sigma;1,\sigma\rangle
  -\frac{3}{\sqrt{55}}|5,4\sigma;1,0\rangle
  +\frac{3}{\sqrt{11}}|5,5\sigma;1,-\sigma\rangle
  \nonumber\\
  &=&
  \sigma\left[
    \sqrt{\frac{2}{165}}\hat{f}^\dagger_{3\sigma,\sigma}\hat{f}^\dagger_{0,\sigma}
    +\frac{1}{\sqrt{165}}\hat{f}^\dagger_{2\sigma,\sigma}\hat{f}^\dagger_{\sigma,\sigma}
    -\frac{3}{\sqrt{110}}\sum_{\sigma'=\pm}\hat{f}^\dagger_{3\sigma,\sigma'}\hat{f}^\dagger_{\sigma,-\sigma'}
    +\frac{3}{\sqrt{11}}\hat{f}^\dagger_{3\sigma,-\sigma}\hat{f}^\dagger_{2\sigma,-\sigma}
    \right]|0\rangle,
  \label{eq:Jz4}\\
  |M_J=3\sigma\rangle
  &=&
  \sqrt{\frac{3}{55}}|5,2\sigma;1,\sigma\rangle
  -\frac{4}{\sqrt{55}}|5,3\sigma;1,0\rangle
  +\frac{6}{\sqrt{55}}|5,4\sigma;1,-\sigma\rangle
  \nonumber\\
  &=&\frac{\sigma}{\sqrt{55}}\left[
    \hat{f}^\dagger_{3\sigma,\sigma}\hat{f}^\dagger_{-\sigma,\sigma}
    +\sqrt{2}\hat{f}^\dagger_{2\sigma,\sigma}\hat{f}^\dagger_{0,\sigma}
    -\frac{4}{\sqrt{3}}\sum_{\sigma'=\pm}\left(\hat{f}^\dagger_{3\sigma,\sigma'}\hat{f}^\dagger_{0,-\sigma'}
    +\frac{1}{\sqrt{2}}\hat{f}^\dagger_{2\sigma,\sigma'}\hat{f}^\dagger_{\sigma,-\sigma'}\right)
    +6\hat{f}^\dagger_{3\sigma,-\sigma}\hat{f}^\dagger_{\sigma,-\sigma}
    \right]|0\rangle,\ \ \ 
  \label{eq:Jz3}\\
  |M_J=2\sigma\rangle
  &=&
  \sqrt{\frac{6}{55}}|5,\sigma;1,\sigma\rangle
  -\sqrt{\frac{21}{55}}|5,2\sigma;1,0\rangle
  +2\sqrt{\frac{7}{55}}|5,3\sigma;1,-\sigma\rangle
  \nonumber\\
  &=&\frac{\sigma}{\sqrt{11}}\left[
    \frac{1}{\sqrt{7}}\hat{f}^\dagger_{3\sigma,\sigma}\hat{f}^\dagger_{-2\sigma,\sigma}
    +\sqrt{\frac{27}{35}}\hat{f}^\dagger_{2\sigma,\sigma}\hat{f}^\dagger_{-\sigma,\sigma}
    +\sqrt{\frac{2}{7}}\hat{f}^\dagger_{\sigma,\sigma}\hat{f}^\dagger_{0,\sigma}
    -\sqrt{\frac{7}{5}}\sum_{\sigma'=\pm}\left(\frac{1}{\sqrt{2}}\hat{f}^\dagger_{3\sigma,\sigma'}\hat{f}^\dagger_{-\sigma,-\sigma'}
    +\hat{f}^\dagger_{2\sigma,\sigma'}\hat{f}^\dagger_{0,-\sigma'}\right)
    \right.\nonumber\\
    &&\left.\ \ \ 
    +2\sqrt{\frac{14}{15}}\hat{f}^\dagger_{3\sigma,-\sigma}\hat{f}^\dagger_{0,-\sigma}
    +2\sqrt{\frac{7}{15}}\hat{f}^\dagger_{2\sigma,-\sigma}\hat{f}^\dagger_{\sigma,-\sigma}
    \right]|0\rangle,
  \label{eq:Jz2}\\
  |M_J=\sigma\rangle
  &=&
  \sqrt{\frac{2}{11}}|5,0;1,\sigma\rangle
  -2\sqrt{\frac{6}{55}}|5,\sigma;1,0\rangle
  +\sqrt{\frac{21}{55}}|5,2\sigma;1,-\sigma\rangle
  \nonumber\\
  &=&\frac{\sigma}{\sqrt{11}}\left[
    \frac{1}{\sqrt{21}}\hat{f}^\dagger_{3\sigma,\sigma}\hat{f}^\dagger_{-3\sigma,\sigma}
    +\frac{4}{\sqrt{21}}\hat{f}^\dagger_{2\sigma,\sigma}\hat{f}^\dagger_{-2\sigma,\sigma}
    +\frac{5}{\sqrt{21}}\hat{f}^\dagger_{\sigma,\sigma}\hat{f}^\dagger_{-\sigma,\sigma}
    -\sqrt{\frac{2}{7}}\sum_{\sigma'=\pm}\hat{f}^\dagger_{3\sigma,\sigma'}\hat{f}^\dagger_{-2\sigma,-\sigma'}
    \right.\nonumber\\
    &&\left.\ \ \ 
    -3\sqrt{\frac{6}{35}}\sum_{\sigma'=\pm}\hat{f}^\dagger_{2\sigma,\sigma'}\hat{f}^\dagger_{-\sigma,-\sigma'}
    -\frac{2}{\sqrt{7}}\sum_{\sigma'=\pm}\hat{f}^\dagger_{\sigma,\sigma'}\hat{f}^\dagger_{0,-\sigma'}
    +\sqrt{\frac{7}{5}}\hat{f}^\dagger_{3\sigma,-\sigma}\hat{f}^\dagger_{-\sigma,-\sigma}
    +\sqrt{\frac{14}{5}}\hat{f}^\dagger_{2\sigma,-\sigma}\hat{f}^\dagger_{0,-\sigma}
    \right]|0\rangle,\ \ \ \ \ 
  \label{eq:Jz1}\\
  |M_J=0\rangle
  &=&
  \sqrt{\frac{3}{11}}|5,-1;1,1\rangle
  -\sqrt{\frac{5}{11}}|5,0;1,0\rangle
  +\sqrt{\frac{3}{11}}|5,1;1,-1\rangle
  \nonumber\\
  &=&\sqrt{\frac{5}{77}}\sum_{\sigma=\pm}\left[\sigma\left(
    \frac{1}{\sqrt{2}}\hat{f}^\dagger_{3\sigma,-\sigma}\hat{f}^\dagger_{-2\sigma,-\sigma}
    +\sqrt{\frac{27}{10}}\hat{f}^\dagger_{2\sigma,-\sigma}\hat{f}^\dagger_{-\sigma,-\sigma}
    +\hat{f}^\dagger_{\sigma,-\sigma}\hat{f}^\dagger_{0,-\sigma}\right)
    \right.\nonumber\\
    &&\left.\ \ \ \ \ \ \ \ \ 
    -\frac{1}{2\sqrt{3}}\hat{f}^\dagger_{3,\sigma}\hat{f}^\dagger_{-3,-\sigma}
    -\frac{2}{\sqrt{3}}\hat{f}^\dagger_{2,\sigma}\hat{f}^\dagger_{-2,-\sigma}
    -\frac{5}{2\sqrt{3}}\hat{f}^\dagger_{1,\sigma}\hat{f}^\dagger_{-1,-\sigma}
    \right]|0\rangle.
  \label{eq:Jz0}
\end{eqnarray}
  \label{eq:Jz43210}
\end{subequations}
\end{widetext}


\begin{thebibliography}{9}
\bibitem{phase}
The notation of the phase in the $q$ term of the Hamiltonian has been changed from $\phi_{\bm{r},\bm{r}'}$ in the previous Letter Ref.\onlinecite{onoda:09} to $2\phi_{\bm{r},\bm{r}'}$ in the present paper, but the physics is totally the same.
\end{thebibliography}
\end{document}